%% file: knot-matching.tex
\documentclass[aoas,preprint]{imsart}

\RequirePackage[OT1]{fontenc}
\RequirePackage{amsthm,amsmath}
\RequirePackage{natbib}
\RequirePackage[colorlinks,citecolor=blue,urlcolor=blue]{hyperref}

\usepackage[utf8]{inputenc} 
\usepackage[T1]{fontenc}    
\usepackage{hyperref}       
\usepackage{url}   	         
\usepackage{booktabs}       
\usepackage{amsfonts}       
\usepackage{nicefrac}       
\usepackage{microtype}      
\usepackage{textcomp,mathrsfs}
\usepackage{graphicx,amssymb,amsmath,amsthm,mathrsfs}
\usepackage{multirow,makeidx,algorithmic,algorithm}
\usepackage{latexsym,graphicx,amssymb,amsmath,amsthm,mathrsfs}
\usepackage{mathrsfs}
\usepackage{booktabs,fancyhdr}
\usepackage[toc,page]{appendix}

\input{basic-math-macros}

\arxiv{arXiv:0000.0000}

\startlocaldefs
\numberwithin{equation}{section}
\theoremstyle{plain}

\endlocaldefs

\begin{document}

\begin{frontmatter}
\title{Sequential Decision Model for Inference and Prediction on Non-Uniform Hypergraphs with Application to Knot Matching from Computational Forestry}
\runtitle{Knot Matching}

\begin{aug}
\author{\fnms{Seong-Hwan Jun}\thanksref{m1}\ead[label=e1]{seong.jun@stat.ubc.ca}},
\author{\fnms{Samuel W.K. Wong}\thanksref{m2}\ead[label=e2]{swkwong@stat.ufl.edu}},
\author{\fnms{James V. Zidek}\thanksref{m1}\ead[label=e3]{jim@stat.ubc.ca}}
\and
\author{\fnms{Alexandre Bouchard-C\^ot\'e}\thanksref{m1}\ead[label=e4]{bouchard@stat.ubc.ca}}

\runauthor{Jun, Wong, Zidek, and Bouchard-C\^ot\'e}

\affiliation{University of British
Columbia\thanksmark{m1}}
\affiliation{University of Florida\thanksmark{m2}}


\end{aug}

%
%

\input{0abstract}

\end{frontmatter}

\input{1introduction}
\input{2background}
\input{3model}

\input{4smc}

\input{simulation}
\input{5data-analysis}

\input{6conclusion}
\subsubsection*{Acknowledgement}

This work was supported by FPInnovations and a CRD grant from the Natural Sciences and Engineering Research Council of Canada. The authors thank FPInnovations, in particular, Zarin Pirouz, Bruce Lehmann, and Alex Precosky for collection and processing of the data.

\bibliography{knot-matching}
\bibliographystyle{imsart-nameyear}

\end{document}

%% file: basic-math-macros.tex
\newcounter{xxx}
\DeclareMathOperator{\argmax}{argmax}

\def\R{{\mathbb R}}        

\def\E{{\mathbb E}}        
\def\1{{\mathbf 1}}        

\newtheorem{theorem}{Theorem}

\newtheorem{definition}[theorem]{Definition}

%% file: 0abstract.tex

\begin{abstract}
In this paper, we consider the \emph{knot matching} problem arising in computational forestry. The knot matching problem is an important problem that needs to be solved to advance the state of the art in automatic strength prediction of lumber. We show that this problem can be formulated as a quadripartite matching problem and develop a sequential decision model that admits efficient parameter estimation along with a sequential Monte Carlo sampler on graph matching that can be utilized for rapid sampling of graph matching. We demonstrate the effectiveness of our methods on 30 manually annotated boards and present findings from various simulation studies to provide further evidence supporting the efficacy of our methods.
\end{abstract}

%% file: 1introduction.tex

\section{Introduction}

Wood processed in mills to produce sawn lumber for use in construction is assigned into grades, according to established rules and standards [\cite{green94}].  The grading process for a piece of lumber involves identifying its visual characteristics and assessing its strength in a non-destructive manner.  The ``knot'', formed by a branch or limb during growth of the tree, is an important class of visual characteristics that affects both the aesthetic quality as well as the strength of wood.  For individual pieces of lumber, previous studies have shown a strong relationship between the size of its knots and its strength when loaded to failure [see for e.g., \cite{castera1996prevision,Hietaniemi2011}]. Therefore, knots have an important role in determining the grade of a piece.

Many modern mills utilize machine vision systems to automate the production process.  Scanning systems incorporating lasers and cameras are used to detect the visual characteristics for quality control and for grading [\cite{brannstrom2009impact,hietaniemi2014real}].  Although images from such systems could be analyzed to provide detailed information about every knot on the piece, current grading rules define standards and size limits on individual knots (or knot clusters) only.  Therefore, much of the potential in the use of these systems to improve lumber strength prediction has yet to be realized.  The strength-reducing effects of different knots on the piece may work together, and jointly modeling their effects may permit more accurate predictions of the ultimate strength of lumber, compared to models based on a single knot alone. Towards this objective, fast and accurate algorithms for detecting and identifying all knots from surface scans of boards are needed.   On that basis, a new strength prediction model can be developed from the complete knot data; uncertainties in the grading and identification can be captured in a probabilistic prediction framework [see for example \citet{wong2016quantifying}].

In this paper, we consider surface scans of lumber pieces that provide images of its four long sides.  In processing these images, two main tasks must be performed to detect and identify knots.  The first task is knot face recognition from the images;  this task belongs mainly in the realm of computer vision as it shares similarities with the object recognition problem.   The second task, which is the focus of this paper, is automatically identifying which of the detected knot faces on the different sides are from the same tree branch; we refer to this problem as ``knot matching''. Note that \emph{knot} refers to the three dimensional convex body that is to be reconstructed from the \emph{knot faces} that are observed on the surfaces of a piece from scanning technologies.  By combining information from four-sided scans, the three-dimensional structure of the wood fibers can be characterized, which is important for strength prediction [\cite{olsson2013prediction}].

Formally, we shall represent a piece of lumber by a quadripartite graph with each of its surfaces forming a partition, with the knot faces as the nodes of the graph.  Knot matching can thus be formulated as a quadripartite matching problem on a non-uniform hypergraph.  We propose a sequential decision model to build a matching, where each decision is modeled via a local multinomial regression.  This class of models is commonly used in other application areas (for example, part-of-speech tagging in natural language processing [see for e.g., \citet{Berg2010Painless}]).  This approach allows inference for the model parameters via maximum likelihood or \textit{maximum a posteriori} estimation to be performed using standard techniques, when given a sample of boards with known matchings (e.g., manually matched by a human).  We then develop a sequential Monte Carlo (SMC) sampler for sampling knot matchings given the estimated parameters. The SMC sampler draws a population of particles from the space of matchings, and thus also serves to estimate uncertainty in the unknown matching when applied to a future piece of lumber. We show that our SMC sampler is fast and thus permits online application for grading in lumber mills. 


Thus we anticipate that our contributions to this applied problem will enhance the sawn lumber production process in two important ways. First, each individual knot can be better assessed by capturing information from all of its visible faces much like a traditional human grader would, thereby increasing the effectiveness of automated grading without sacrificing speed and efficiency.  Second, accurate automatic knot matchings will provide an important component of the necessary data for future refinement of lumber strength prediction models based on visual characteristics.


The paper is organized as follows. In Section~\ref{sec:background-data}, we describe the process that generates the data for knot matching.  In Section~\ref{sec:background-matching} we introduce the graph theory notion relevant for problem formulation and provide an overview of probabilistic graph matching and related work.  In Section~\ref{sec:sequential-deicision-model}, we develop a sequential decision model for constructing a matching and in Section~\ref{sec:inference}, we show that this model admits efficient parameter inference.  In Section~\ref{sec:prediction}, we develop an SMC sampler for drawing samples from the distribution of matchings defined on a non-uniform quadripartite hypergraph and how it can be utilized for prediction. In Section~\ref{sec:simulation} we present a procedure to simulate realistic knot matching data for evaluating our model and SMC sampler. We present experimental results on simulated and real data in Section~\ref{sec:application}, and conclude the paper with a brief discussion in Section~\ref{sec:conclusion}.

%% file: 2background.tex

\section{Data for knot matching}
\label{sec:background-data}

The images necessary for the development of our application are generated for a piece of lumber, as shown in Figure~\ref{fig:lumberD4006}.  High-definition cameras are installed to capture high-definition images of the four surfaces as it moves on a carrier (i.e., each piece is taken along conveyor belt through the scanning station).  Note that each piece of lumber has six sides but the two ends are typically ignored, as controlled sawing leaves no knot faces to appear on the end sides.  The first processing task is to identify the \emph{knot faces} in the images.  This task is much like object detection and localization (e.g., face detection from a photo).  For this purpose we use an internally developed knot detection algorithm that outputs data on the location and size of the knot faces. The second task, and the focus of this paper, is to identify which of the knot faces on the different surfaces belong to the same knot (i.e.,~from the same tree branch).  In this section, we provide the details of the relevant data for the knot matching problem.



\subsection{Lumber and knot representation}

\begin{figure}[t!]
	\includegraphics[width=0.95\textwidth]{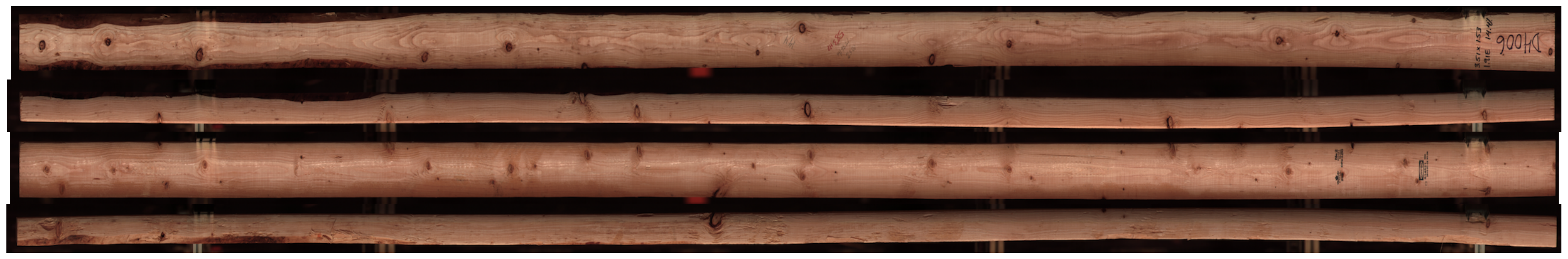}
	\caption{Sample lumber used in the real data analysis. Four sides of the boards with the wide surfaces shown on the first and the third rows and the narrow surfaces shown on the second and the last rows.}
	\label{fig:lumberD4006}
\end{figure}

\begin{figure}[t!]
	\includegraphics[width=0.95\textwidth]{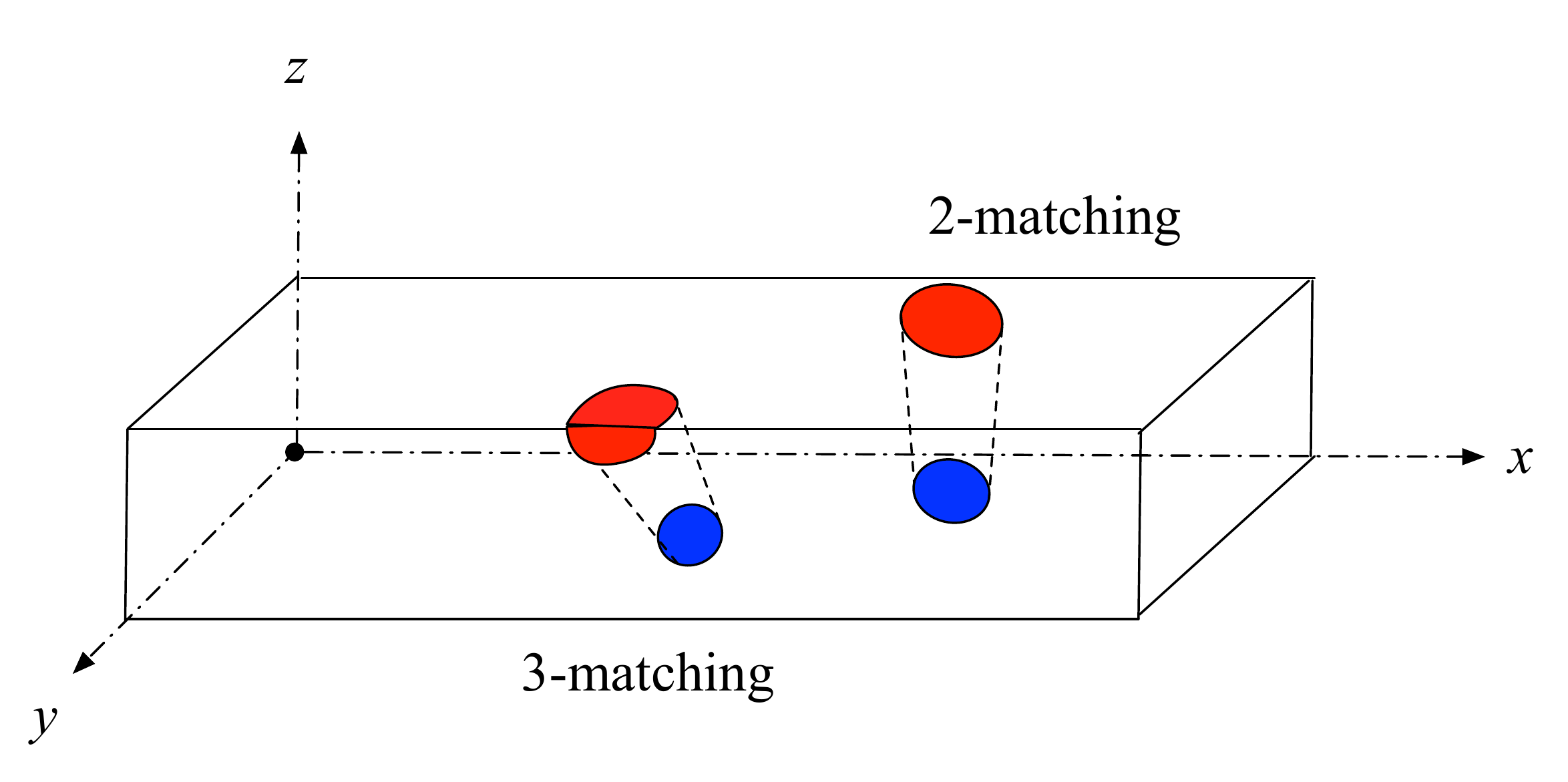}
	\caption{3-dimensional view of the lumber (not to scale). An illustration of 2-matching and 3-matching are provided.}
	\label{fig:board-3d}
\end{figure}

The raw images of the surfaces as shown in Figure~\ref{fig:lumberD4006} are first processed by an internally developed knot detection algorithm.  Knots are often modelled as elliptical cones [\cite{guindos2013three}], and hence, our implementation of the knot detection algorithm models each knot face as an ellipse. We view each piece of lumber as a 3-dimensional object, positioned in a standard 3-dimensional Euclidean space as shown in Figure~\ref{fig:board-3d}, with the $x$, $y$ and $z$-axes representing length, width, and height respectively.  In this fashion, the two `wide' surfaces in Figure~\ref{fig:lumber24} (a) and (c) are parallel to the $x$-$y$ plane, while the two `narrow' surfaces in Figure~\ref{fig:lumber24} (b) and (d) are parallel to the $x$-$z$ plane.  


For each knot face on each surface, the knot detector outputs the 3-dimensional coordinate, $(x, y, z)$, indicating the position of the center of the knot face.  It also outputs the axes of the fitted ellipse on the knot face, denoted by $(a, b)$ where $a$ is the length of the axes along the $x$-axis and $b$ is the length of the axes along the $y$-axis for the two `wide' surfaces and $z$-axis for the two `narrow' surfaces. Additionally, we have the rotation angle of the fitted ellipse, denoted $\alpha$. In summary, each knot face is represented by the 6-tuple $(p, x, y, z, a, b, \alpha)$, where $p$ denotes the index of the surface (i.e., partition).



\subsection{Choice of covariates}

For knot faces on distinct surfaces, we can compute a vector of associated covariates to assess whether the knot faces belong to the same branch, and hence should be matched together.  Covariates that are useful predictors would help distinguish matches from non-matches.  For each pair of knot faces $u$ and $v$ on distinct surfaces we considered the following covariates:
\begin{itemize}
	\item Both $u$ and $v$ appear on a wide surface:  We compute the distance between $u$ and $v$.  We observed from our data the most common occurrence among the matched knots is of this type.  Knot faces at shorter distances apart are more likely to be matches.
	\item One of $u$ or $v$ appears on a narrow surface:  This covariate resembles the one above in that we compute the Euclidean distance between $u$ and $v$. We found that differentiating the two cases helped improve the performance of the matching method that we will describe in the subsequent sections.
	\item Comparison of sizes: To assess the size difference between knot faces, we compute the areas of their fitted ellipses and compute the absolute difference $|u_{area} - v_{area}|$.  Knot faces belonging to the same branch are expected to have a smaller difference in sizes. 
\end{itemize}

For a triplet of knot faces $u, v, w$, we consider the following covariate, which is a slight modification of the ones listed above to accommodate 3-matching:
\begin{itemize}
	\item Maximum and minimum distances: We compute the Euclidean distance between each pair of knot faces and extract the maximum and the minimum pairwise distances as covariates. Recall that the knot faces represent a surface of a convex body that appear when an elliptical cone is sliced. Therefore, two of the knot faces must share an axis as shown in Figure~\ref{fig:board-3d}. However, we found that inaccuracies during the knot detection stage can potentially capture two knot faces that share an axis and appear separated. This error is of a reasonable size and we found that computing the distance between the nearest knot faces instead is a useful approximation that leads to good empirical performance. The maximum distance is analogous to the distance covariate computed above for a pair of knot faces.
	\item Comparison of sizes:  To adapt the size covariate above, we sum the area of the fitted ellipses of the two closest knot faces and take the absolute difference with the area of the remaining knot face.
\end{itemize}

These covariates are incorporated in the matching model developed in Section~\ref{sec:sequential-deicision-model}. We estimate the parameters associated with each of these covariates from a sample of boards where the correct knot matching is known.






\subsection{Matching by a human grader}

\begin{figure}[t!]
\begin{tabular}{c}
	\includegraphics[width=0.95\textwidth]{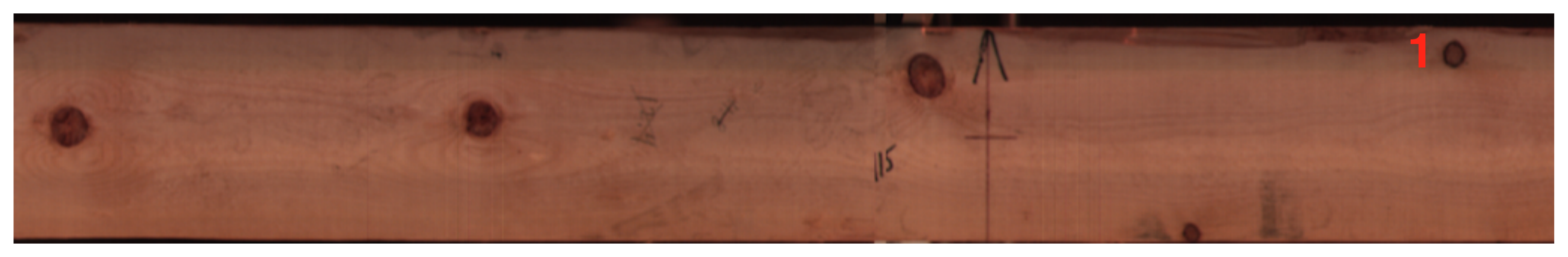} \\
	\includegraphics[width=0.95\textwidth]{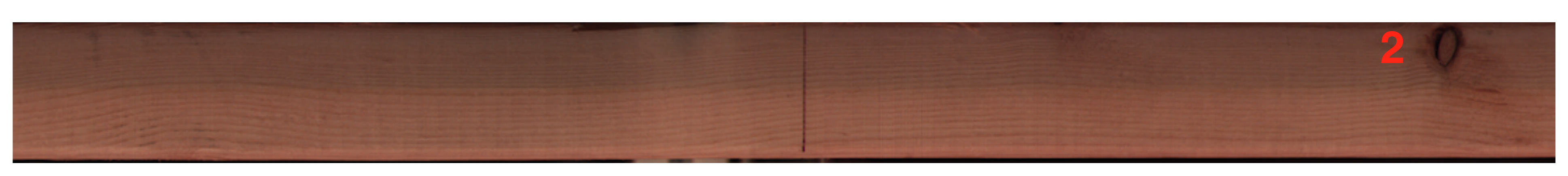} \\
	\includegraphics[width=0.95\textwidth]{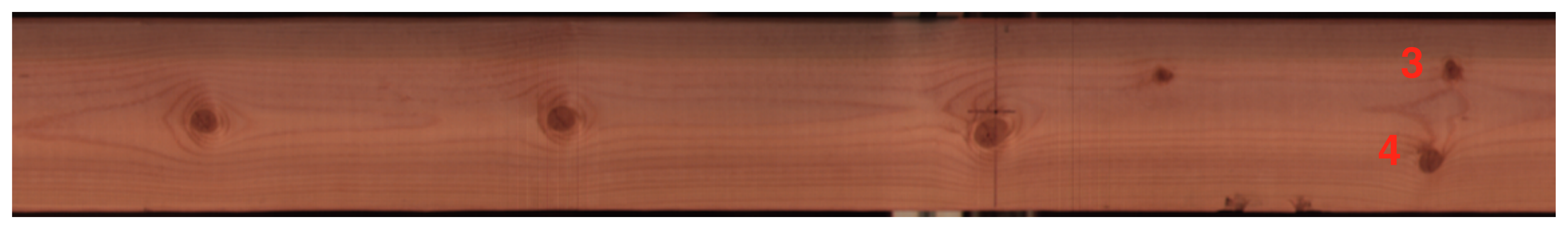} \\
	\includegraphics[width=0.95\textwidth]{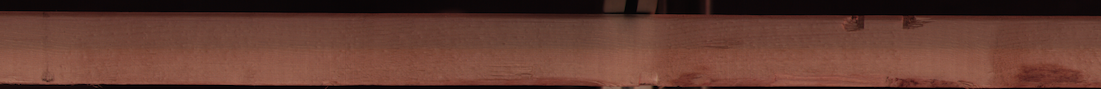} \\
\end{tabular}
\caption{A closer look at a segment of a plank. The matching for knot faces labelled $1, 2, 3, 4$, produced by the human grader is $\{ \{1, 4\}, \{2, 3\} \}$.}
\label{fig:lumber24}
\end{figure}

For a well trained person, matching the knot faces is an easy task (although time consuming).  That human grader would examine a piece of lumber from end-to-end, using the visual characteristics of the knot faces to determine the matches.  For example, note the knots labelled 1 to 4 in red, on Figure~\ref{fig:lumber24}.  The grader is able to determine the correct matching after careful examination:  knot face 1 and knot face 4 belong to the same branch and knot face 2 and knot face 3 belong to the same branch.  However, there are cases that can be difficult even for a human grader and we would like to quantify uncertainty in the matching using probabilities. 

We utilized a human grader to manually annotate our data. For each board, each knot is represented by $(p, x, y, z, a, b, label)$, where $label$ is a unique identifier given to knot faces that stem from a same branch.  The manually annotated matchings will be used to evaluate the performance of our approach.


\section{Overview of graph matching}
\label{sec:background-matching}

We denote a \emph{graph} by $G=(V, E)$ where $V$ is a set of nodes in the graph and $E$ denotes the set of edges.  We shall require several extensions to the basic notion of a graph, which we briefly introduce.  In a \emph{hypergraph}, an edge can contain any number of nodes. A hypergraph is $d$-\emph{uniform} if all of the edges contain $d$ nodes.  A \emph{$K$-partite} hypergraph is denoted by $G=(V_1, V_2, \ldots, V_K, E)$ where $V_1, V_2, \ldots, V_K$ are disjoint sets of nodes referred to as partition sets and $E$ denotes a set of edges such that each edge $e \in E$ may not contain two nodes of the same partition.  The special case with $K=2$ is known as a \emph{bipartite} graph.

In graph theory terms, each surface of the board represents a partition set and the knot faces appearing on the surfaces correspond to the nodes.  We can let the edge set contain any combination of knot faces as long as it does not contain knot faces from the same surface.  A tree branch cannot manifest itself more than once on any one surface. Hence, we have the same restriction as in the $K$-partite graphs, where no edge can contain nodes from the same partition.  Formally then, we view a piece of lumber as a complete (non-uniform) $4$-partite hypergraph.




A typical set up for a maximal graph matching problem is that given a graph $G=(V, E)$ and a weight function $w : E \to [0, \infty)$, we wish to find a matching $M \subset E$ such that $\sum_{e \in M'} w(e) \leq \sum_{e \in M} w(e)$ for any other matching $M' \subset E$.  Usually, this requires (i) computing the weight function, (ii) finding high-weight matchings. We provide an overview below and develop our approaches in the following sections.


\subsection{Computing the weight function} A common practice in machine learning is to choose a parametric model for the weight function. One common model is the \emph{Gibbs measure} [\cite{petterson2009exponential,Bouchard2010Variational}]:

\begin{equation}\label{eq:gibbs-measure}
	p(M = m | \theta) = \frac{e^{-w(m; \theta)}}{\sum\limits_{m' \in \mathcal{M}} e^{-w(m; \theta)}},
\end{equation}
which defines a probability distribution over the space of matchings. Here, we let $M$ denote a random matching and $m$, a realized matching, with $\mathcal{M}$ denoting the sample space of matchings for a given graph. The parameters are denoted by $\theta \in \mathbb{R}^d$ and $w(m; \theta)$ denotes the weight of the given matching $m$. Note that we have overloaded the notation for $w$ to be defined on the edges as well as matchings. 

Given $\theta$, we can use Equation~\ref{eq:gibbs-measure} to compute the likelihood for any matching.   A linear function $w(m; \theta) = -\sum_{e \in m} \theta^T \phi(e)$ is commonly adopted as the weight function, where $\phi(e)$ denotes the covariates extracted from edge $e \in m$.  This model is preferred in the machine learning community because of its \emph{exchangeability}, i.e., the order in which the edges in the matching are observed does not affect the probability of a given matching.

However, inference for the parameters based on the model in Equation~\ref{eq:gibbs-measure} can be quite challenging due to the normalization constant, whose computation requires enumeration over the space of matchings.  Given a sample of known graph matchings, computing the gradient of Equation~\ref{eq:gibbs-measure} is inefficient and hence, rules out gradient-based procedures for optimization of the parameters (for example, to find the maximum likelihood estimate) [\cite{petterson2009exponential}]. For the same reason, it is difficult to sample from the posterior distribution of the parameters given data.  Standard Bayesian inference methods such as MCMC would encounter the so-called \emph{doubly intractable} problem, where one would have to compute the normalization constant at each iteration of the MCMC to evaluate the Metropolis-Hastings acceptance probability [\cite{moller2006efficient}].


We propose to replace the Gibbs measure by a sequential decision model for this application. The sequential decision model is motivated by the process used by a human grader manually matching the knots. A human grader inspecting a piece of lumber proceeds by sequentially examining one knot face at a time, from one end of the piece to the other.  The grader relies on the approximate distance, size, and shape of the knot faces to arrive at the current matching following this sequential procedure. We thus expect that modelling this process is useful as a step towards developing an automatic matching algorithm. We shall see in Section~\ref{sec:inference} that our model admits efficient parameter inference.

A similar model restricted to sampling bipartite matching has been proposed in [\cite{Volkovs2012}] but it does not address the problem of parameter inference. Matching on graph has been tackled using an approximate counting schemes such as [\cite{jerrum2004polynomial}]. However, the focus of such methods is restricted to bipartite graphs and developing an MCMC sampler for $K$-partite hypergraph does not appear to be trivial.


\subsection{Searching for matchings}
Finding the maximal matching is a combinatorial optimization problem that has been extensively studied in the graph theory community [see e.g., \cite{kuhn1955hungarian,bondy1976graph,papadimitriou1982combinatorial}]. In particular, the bipartite matching problem has received much attention due to its wide array of applicability to disciplines such as computer vision, computational biology, and information retrieval among others [\cite{holmes2001pairwise,lunter2005statistical,cao2007learning,caetano2009learning}].  In statistics, the problem of bipartite matching has been applied to the design of experiments when a pair of similar subjects need to be matched, and to causal inference where the goal is to match similar observational units [see e.g., \cite{hansen2004full,lu2004optimal}].  Given the weight function, there are deterministic algorithms such as the Hungarian algorithm that can find a maximal matching in polynomial time for bipartite graphs [\cite{kuhn1955hungarian}], however such algorithms are unavailable for general $K$-partite graphs.


A key development in the sequential Monte Carlo literature that we depend on in this paper appears in \citet{del2006sequential}, which allows SMC to be used for inference for general state spaces, coupled with theoretical developments in \citet{Wang2015Bayesian}, which establishes necessary conditions for using SMC for combinatorial spaces. Thus SMC has become a suitable choice for sampling graph matching.  Building on these previous works, we proposed an SMC sampler for graph matching in \cite{jun2017sequential}. Our work in \cite{jun2017sequential} improved on the previous work using SMC for graph matching proposed in \cite{suh2012graph} in three ways: 1) we consider the problem of sampling graph matching on a hypergraph rather than restricting to bipartite graphs, 2) we consider the parameter estimation problem, and 3) we address the overcounting problem that can arise when using SMC for sampling from the combinatorial spaces. Further building on that work, the focus of this paper is on describing the details of the SMC sampler and relevant methodologies for tackling the specific challenges of the knot matching problem. In particular, we have completed the analysis of the knot matching data by considering 3-matchings as well as carrying out the performance analysis on a larger data set.

%% file: 3model.tex

\section{Sequential Decision Model for $K$-partite Hypergraph Matching}
\label{sec:sequential-deicision-model}

In this section, we formalize the idea of modelling the sequential knot matching procedure. The exposition in this section is an improved version of our previous work [\cite{jun2017sequential}].



\subsection{Sequential Decision Model}

We shall consider a matching to be represented by a sequence of decisions.  That is, nodes are visited one-by-one to decide their set membership, given past decisions. Each decision in the sequence is modelled using multinomial logistic regression.  Recall the Gibbs model in Equation~\ref{eq:gibbs-measure}, where we noted that parameter estimation is difficult in general.  In contrast, our sequential representation admits efficient parameter inference. 


Let $\sigma : \{1, ..., |V|\} \to \{1, ..., |V|\}$ be the sequence in which the nodes are visited where $V$ denotes the set of all nodes of a graph $G=(V, E)$.  For ease of exposition, we will first assume that this permutation is known and fixed; in general, this sequence can be either random or deterministic.  Each node $v_{\sigma(r)} \in V$ for $r = 1, ..., |V|$, makes a decision and the decision made by node $v_{\sigma(r)}$ is to be denoted by $d_{v_{\sigma(r)}}$. The set of decisions available for a node is to be denoted by $\mathcal{D}(v_{\sigma(r)}, m_{r-1})$. Here, we use $m_{r-1}$ to denote the partial matching implied by the sequence of decisions, $\{ d_{v_{\sigma(1)}}, ..., d_{v_{\sigma(r-1)}} \}$, i.e., for each decision sequence, we have a mapping $\{ d_{v_{\sigma(1)}}, ..., d_{v_{\sigma(r-1)}} \} \to m_{r-1}$ where $r-1$ nodes have made decisions. In the rest of the paper, we will often omit $m_{r-1}$ for notational simplicity and just write $\mathcal{D}(v_{\sigma(r)})$. Interpretation for the decision set $\mathcal{D}(v_{\sigma(r)})$ is as a set of candidate edges that $v_{\sigma(r)}$ can be placed into.  To be precise, we can think of making a decision $d$ as first forming a new edge $d' = d \bigcup \{v_{\sigma(r)}\}$ and updating a matching by setting $m_{r} = m_{r-1} \setminus d \bigcup d'$.

The decision set is flexible and can be chosen to suit the problem at hand.  For example, in bipartite matching without any restrictions, the decision candidate for a node $v_{\sigma(r)}$ consists of all unmatched nodes in the partition opposite to $v_{\sigma(r)}$. This can be represented by setting
\begin{equation*}
	\mathcal{D}(v_{\sigma(r)}) := \bigcup\limits_{
	\begin{array}{c}
		u \in V_k : \\
		u \notin m_{i-1} \\ 
		v_{\sigma(r)} \notin V_k
	\end{array}} \{ u \}.
\end{equation*}
The decision set formulation also permits singleton sets where a node is placed into a set by itself, which is achieved by including an empty set in the decision set.  We model each decision by a multinomial regression involving the covariates extracted from an edge $d'$, denoted $\phi(d')$, and the parameter vector $\theta$:
\begin{equation}\label{eq:local-model}
	p( d_{v_{\sigma(r)}} | m_{r-1}, \sigma, \theta) = \frac{\exp \left[ \theta^T \phi(d_{v_{\sigma(r)}} \bigcup \{v_{\sigma(r)}\}) \right]}{\sum\limits_{j=1}^{|\mathcal{D}(v_{\sigma(r)})|} \exp\left[ \theta^T  \phi(d_j \bigcup \{v_{\sigma(r)}\}) \right]},
\end{equation}
i.e.,~each decision has a multinomial distribution with $|\mathcal{D}(v_{\sigma(r)})|$ categories. Note that we are not restricted to local covariates; it is also possible to include global features where $\phi$ is defined on the matching, $\phi(m_{r})$.

Taking the product of the local multinomial probabilities induces the likelihood model as follows,

\begin{equation}\label{eq:sequential-likelihood}
	L(d_{\sigma} | \sigma, \theta ) = \prod\limits_{r=1}^{|V|} p(d_{v_{\sigma(r)}} | m_{r-1}, \sigma, \theta).
\end{equation}

This model is akin to the Plackett-Luce model [\cite{plackett1975analysis,caron2014bayesian}], commonly used for modelling ranking and preferences. One can also express the joint distribution of the decisions and the permutation as,

\begin{equation}\label{eq:sequential-decision-model}
	p(d_{\sigma}, \sigma | \theta) = \prod\limits_{r=1}^{|V|} p(d_{v_{\sigma(r)}} | m_{r-1}, \sigma_{r}, \theta) p(\sigma_{r} | \sigma_{r-1}),
\end{equation}

\noindent where $\sigma_{r}$ is a partial map $\sigma_r : \{1, ..., r\} \to \{1, ..., |V|\}$.

\subsection{Specification of Decision Models for Knot Matching}
\label{sec:decision-model}

In this section, we provide details about the decision models we use for the knot matching problem. For ease of exposition, we begin by providing an example of a decision model for the bipartite matching problem and proceed to describe the specifics of a decision model that we have considered for the knot matching application.

\subsubsection{Bipartite Matching}
\label{sec:decision-model-bm}

\begin{figure}[t!]
	\centering
	\begin{tabular}{cc}
		\includegraphics[scale=0.33]{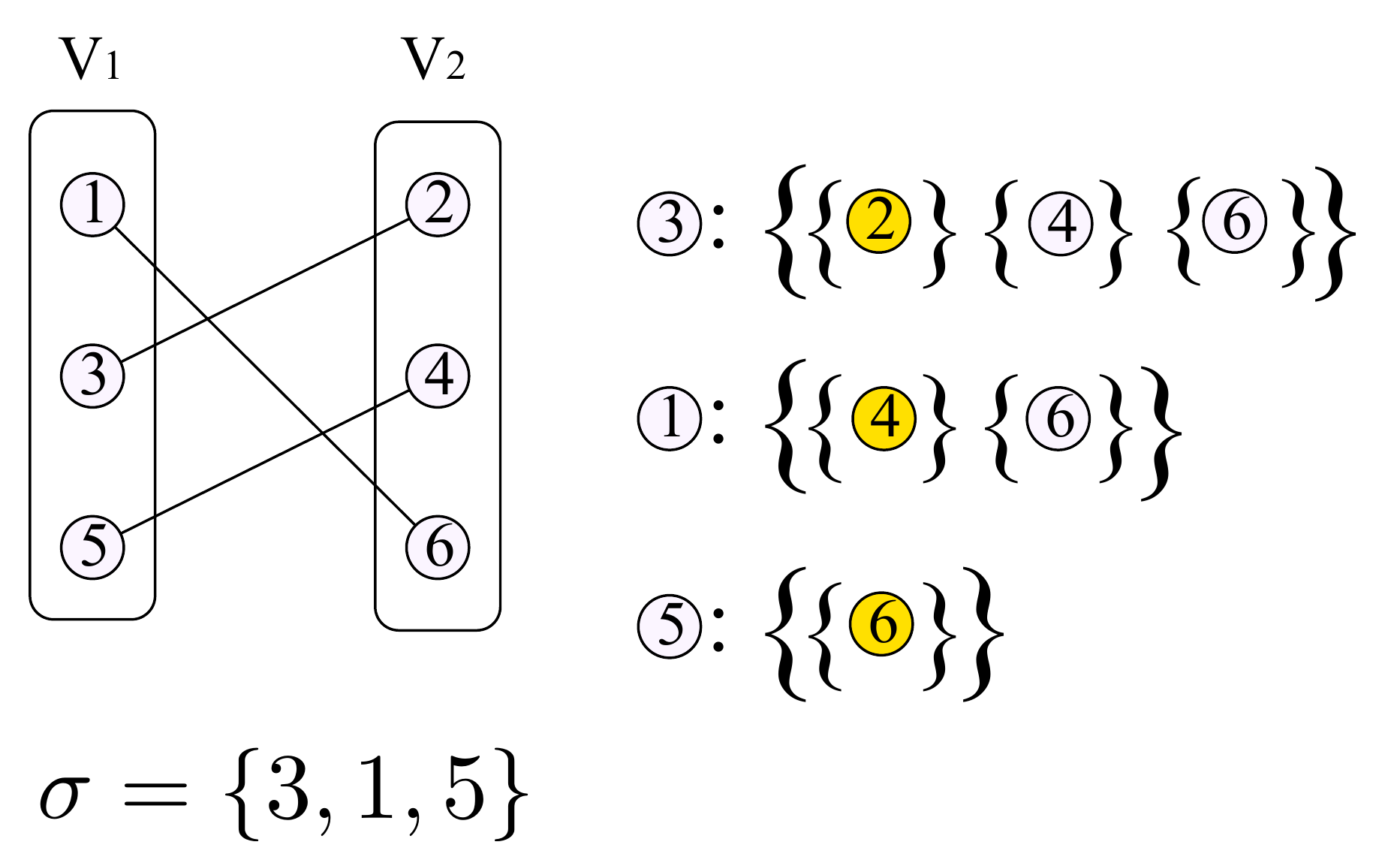} &
		\includegraphics[scale=0.33]{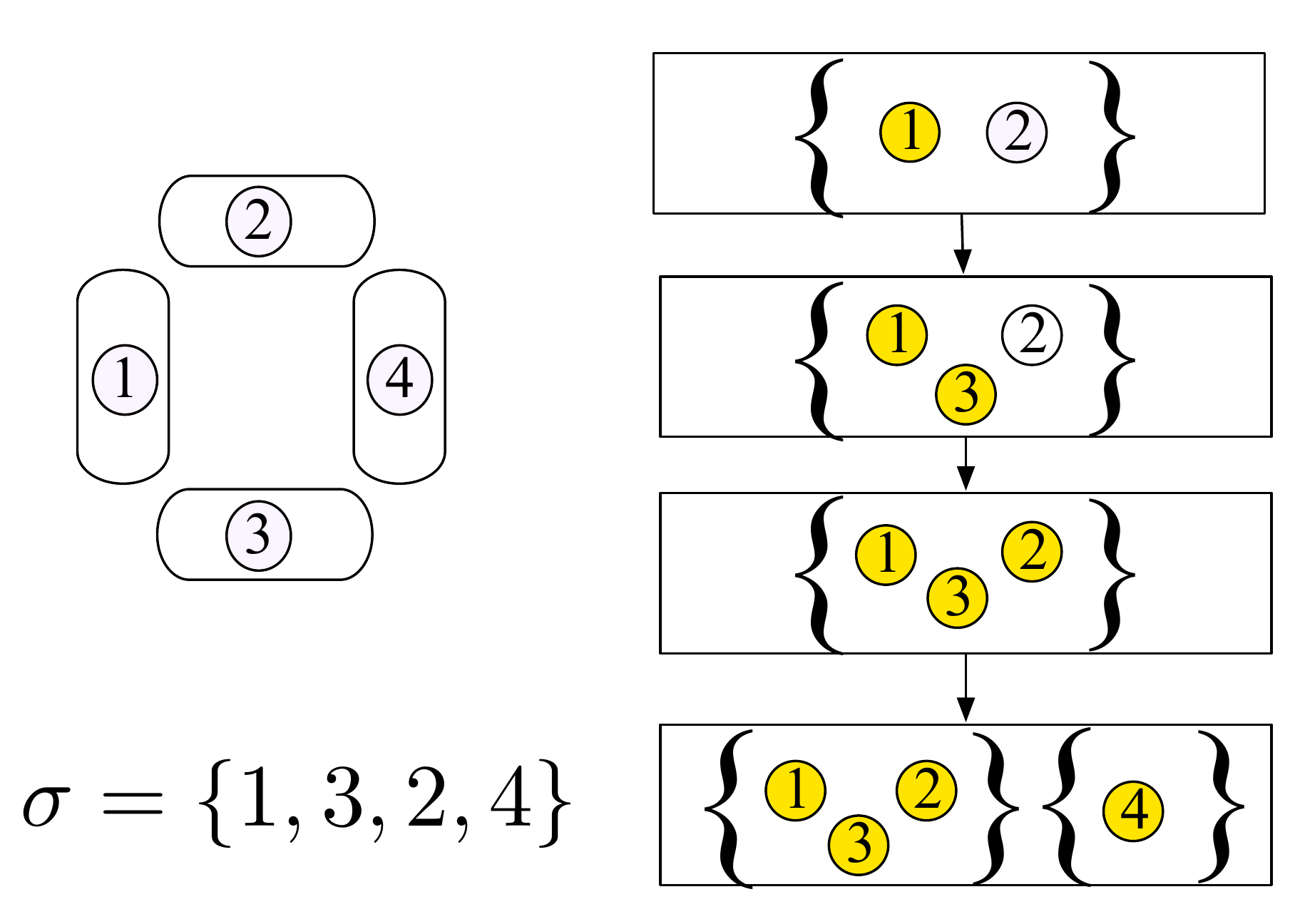} \\
		(a) & (b)
	\end{tabular}
	\caption{(a) Illustration of the decisions for a bipartite matching. For bipartite matching, we only need to visit the nodes in one of the partitions. In this illustration, we have $\sigma = (3, 1, 5)$. The decision set for the nodes are presented with the selected nodes colored in yellow. (b) An illustration of sequential decision model used for knot matching application with each partition containing exactly one knot, labelled from 1 to 4. The rectangles represent matching and curly braces represent edges.  The visited nodes are colored in yellow whereas the nodes that are yet to be visited are unfilled.}
	\label{fig:simple-matching-example}
\end{figure}

Bipartite matching is a special case where it is sufficient to form a  matching by visiting the nodes in only one partition. An illustration of the decision set is shown in Figure~\ref{fig:simple-matching-example} (a). In this illustration, we visit the nodes in $V_1$ with permutation $\sigma = \{ 3, 1, 5\}$. First, the decision set for node $3$ is all of the nodes in $V_2$: $\{ \{2\}, \{4\}, \{6\}\}$. From this set, the node $2$ is chosen (marked as yellow) first. The process continues for node 1 and then node 5 to form the bipartite matching shown in the figure.

\subsubsection{Knot Matching}
\label{sec:decision-model-knot-matching}

\begin{figure}[t!]
	\centering
	\begin{tabular}{c}
		\includegraphics[width=11cm]{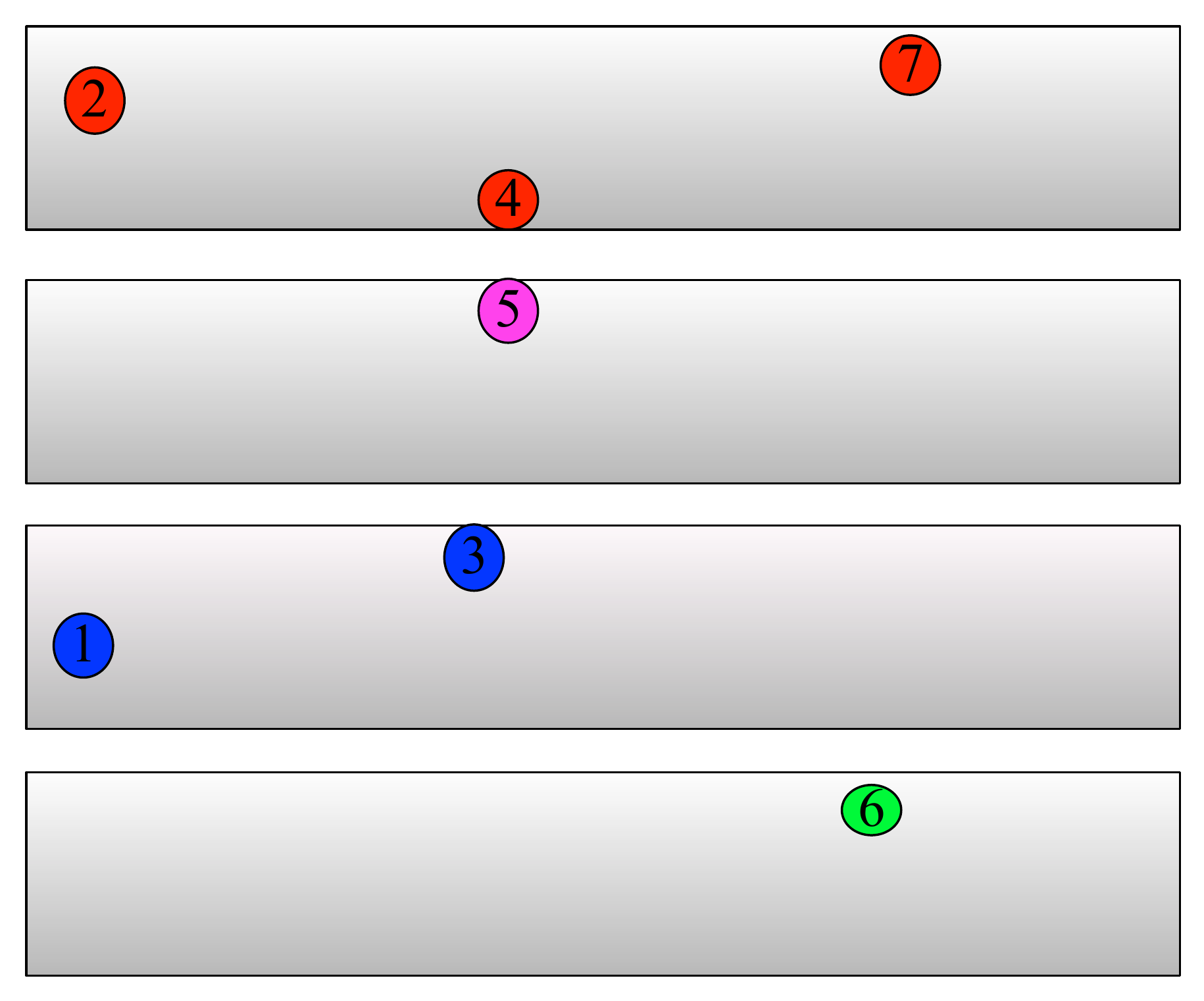} \\
		(a) \\
		\begin{tabular}{c||c||c}
			\includegraphics[width=4.5cm]{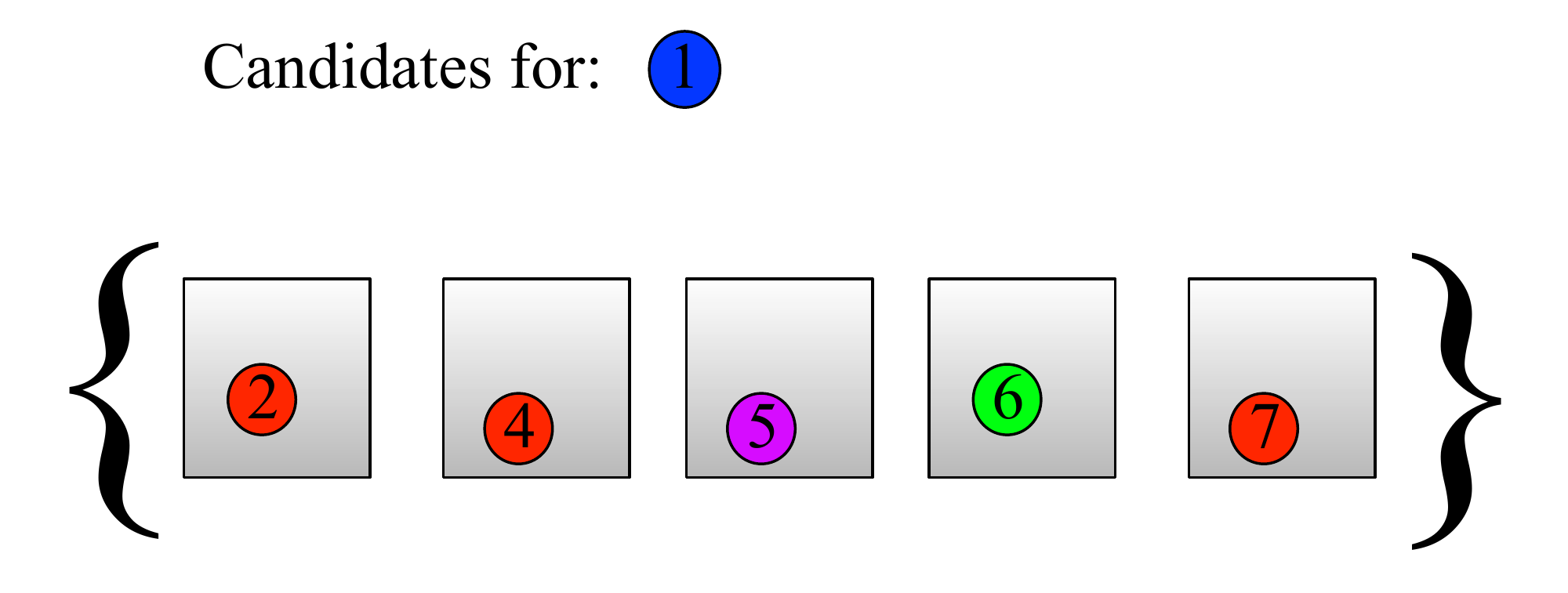} &
			\includegraphics[width=3.5cm]{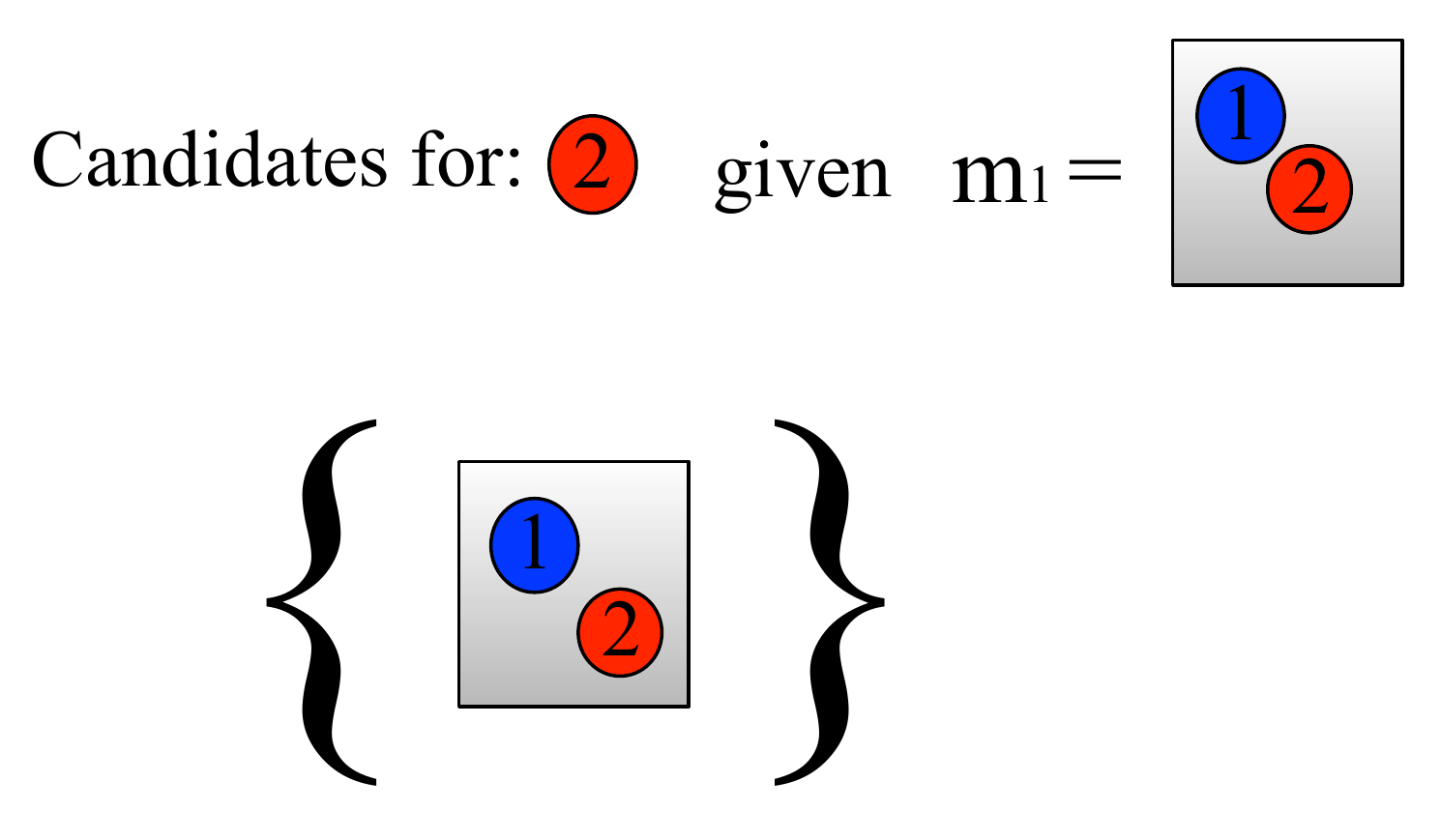} &
			\includegraphics[width=3.5cm]{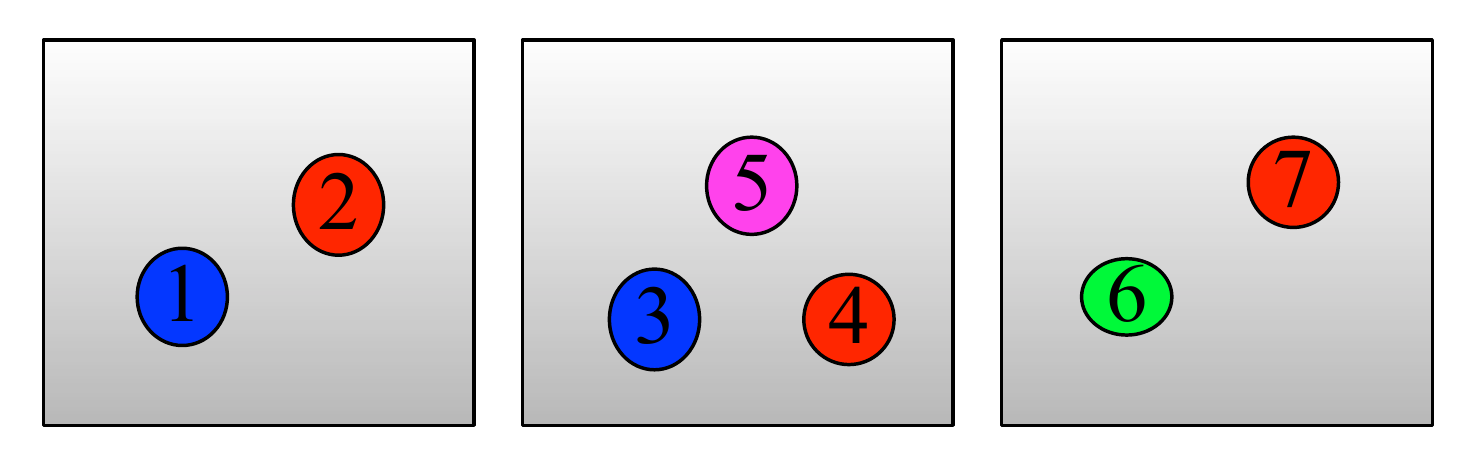} \\
			(b) & (c) & (d)
		\end{tabular} \\
	\end{tabular}
	\caption{(a) 4-partite hypergraph representing a piece of lumber. (b) The decision set for blue node \#1. (c) The decision set for red node \#2 given the decision made by blue node \#1. (d) An example of a final matching. This is a modified version of a similar figure appearing in \cite{jun2017sequential}.}
	\label{fig:4-partite-hypergraph}
\end{figure}

For the knot matching problem, we begin by imposing a restriction on the cardinality of the edges in the matching to be restricted to $\{2, 3\}$. This restriction stems from the fact that $4$-matching is rarely observed in practice. With this restriction in place, a decision set for an uncovered knot $v$ can be formulated to include any knot face on a different surface from $v$ as well as any edge whose cardinality is $2$ and does not contain a knot face from the same partition as $v$. If $v$ is already covered (belongs to an edge), we formulate the decision set with only an empty set, equivalent to allowing no decisions. We have provided an illustration of a knot matching problem in Figure~\ref{fig:4-partite-hypergraph} (a) and illustrative  decision sets in Figure~\ref{fig:4-partite-hypergraph} (b) and (c), with $\sigma=\{1, 2, 3, ..., 7\}$, pre-specified. In Figure~\ref{fig:4-partite-hypergraph} (b), we have $m_0 = \emptyset$, and hence, it considers all of the nodes in the graph with different colors as candidates. In Figure~\ref{fig:4-partite-hypergraph} (c), we note that the red node labelled \#2 is already contained in an edge and hence, the only decision available is an empty set (for illustration purposes, we have indicated the edge that contains it in the decision set). Finally, an example of a final matching state is provided in Figure~\ref{fig:4-partite-hypergraph} (d).

Note that this decision model allows a singleton set as a by-product. For example, consider the case with one node in each partition shown in Figure~\ref{fig:simple-matching-example} (b). Given a permutation $\sigma = (1, 3, 2, 4)$, the decision set for node 1 is $\{2, 3, 4\}$. Suppose it matches with node 2. Node 3 is presented with decision set $\{\{1, 2\}, \{4\}\}$. Suppose it decides to form $\{1, 2, 3\}$. Next, we note that node 2 is already covered, so it is presented with an empty set as the only decision. Then, when we visit node 4, the only decision presented to it is an empty set because the edge $\{1, 2, 3\}$ is already saturated. Therefore, node \#4 forms a singleton set. In practice, a singleton case may arise due to an imperfect knot detection step.

%% file: 4smc.tex
\section{Parameter Estimation via Monte Carlo Expectation Maximization}
\label{sec:inference}

With the model in place, it remains to address the problem of estimating its parameters, which we do in this section.

Assume that we are given a data set of $I$ matchings: $m^1, ..., m^I$. We would like to maximize $p(\theta | m^1, ..., m^I)$. One potential difficulty is that for a given matching $m^i$, there can be multiple paths (i.e., permutation and decision sequences) that lead to $m^i$. For example, consider the bipartite matching $\{ \{1, 6\}, \{3, 2\}, \{5, 4\} \}$ (shown in Figure~\ref{fig:simple-matching-example} (a)). This matching can be attained with $\sigma = (1, 3, 5)$ where $d_{\sigma(1)} = \{ 6 \}$, $d_{\sigma(2)} = \{ 2 \}$, and $d_{\sigma(3)} = \{4\}$ as well as $\sigma = (3, 1, 5)$ where $d_{\sigma(1)} = \{ 2 \}, d_{\sigma(2)} =\{6\}$, and $d_{\sigma(3)} = \{4\}$. 

We view the permutation and the decisions as latent variables and express the complete data likelihood as,
\begin{align}
	\prod\limits_{i=1}^{I} L_c(m^i, \sigma^i, \boldsymbol{d}_{\sigma^i} | \theta) &= \prod\limits_{i=1}^{I} p(m^i | \sigma^i, \boldsymbol{d}_{\sigma^i}) p(\sigma^i, \boldsymbol{d}_{\sigma^i} | \theta) \\
	&= \prod\limits_{i=1}^{I} 1[(\sigma^i, \boldsymbol{d}_{\sigma^i}) \to m^i] \times p(\sigma^i, \boldsymbol{d}_{\sigma^i} | \theta).
\end{align}
Note that $p(\sigma^i, \boldsymbol{d}_{\sigma^i} | \theta)$ is given by Equation~\ref{eq:sequential-decision-model} and the $1[(\sigma^i, \boldsymbol{d}_{\sigma^i}) \to m^i]$ is an indicator function equal to $1$ if and only if $(\sigma^i, \boldsymbol{d}_{\sigma^i})$ maps to $m^i$. The inference can be carried out iteratively using the expectation maximization [\cite{dempster1977maximum}]:
\begin{align*}
	\text{(E step): } & Q(\theta, \theta^t) = \E\left[\log p\left(\theta | \{ m^i, \sigma^i, \boldsymbol{d}_{\sigma^i}\}_{i=1}^{I}\right) \right], \\
	\text{(M step): } & \theta^{t+1} = \argmax_{\theta} Q(\theta, \theta^t).
\end{align*}
where expectation is taken with respect to $(\sigma^i, \boldsymbol{d}_{\sigma^i}) \sim p(\sigma, \boldsymbol{d}_{\sigma} | \theta^t, m^i)$. Note that the posterior can be expressed as,
\begin{align*}
	p(\theta | \{ m^i, \sigma^i, \boldsymbol{d}_{\sigma^i}\}_{i=1}^{I}) &\propto \prod\limits_{i=1}^{I} L_c(m^i, \sigma^i, \boldsymbol{d}_{\sigma^i} | \theta) p(\theta).
\end{align*}

Therefore, the Q function can be expressed as,
\begin{align}
	Q(\theta, \theta^t) &\propto \sum\limits_{i=1}^{I} \E[\log L_c(m^i, \sigma^i, \boldsymbol{d}_{\sigma^i})] + \log p(\theta) \nonumber \\
	&= \sum\limits_{i=1}^{I} \sum\limits_{\sigma^i, \boldsymbol{d}_{\sigma^i}} \log L_c(m^i, \sigma^i, \boldsymbol{d}_{\sigma^i}) p(\sigma^i, \boldsymbol{d}_{\sigma^i} | \theta^t, m^i) + \log p(\theta). \label{eq:Q-function}
\end{align}
We use a Monte Carlo version of EM to approximate the expectation involved in the E-step [\cite{wei1990monte}]. To sample from $p(\sigma, \boldsymbol{d}_{\sigma} | \theta^t, m^i)$, we use sequential Monte Carlo by defining the state space at each iteration of the SMC as $\mathcal{S}_r = \Sigma_r \times \mathcal{D}_r$, where $\Sigma_r$ is the set of all possible permutation sequences of length $r$ and $\mathcal{D}_r$ is the set of all possible decision sequences of length $r$. We will let $\sigma_r \in \Sigma_r$ denote the partial map $\sigma_r : \{1, ..., r\} \to \{1, ..., |V|\}$. The intermediate target distribution for iteration $r$ is, 
\begin{equation*}
	p(\sigma_r^i, \boldsymbol{d}_{\sigma_r^i} | \theta^t, m^i) = \frac{1[(\sigma_r^i, \boldsymbol{d}_{\sigma_r^i}) \in m^i] p(\sigma_r^i, \boldsymbol{d}_{\sigma_r^i} | \theta^t)}{p(m^i | \theta^t)}.
\end{equation*}
We use the notation $(\sigma_r^i, \boldsymbol{d}_{\sigma_r^i}) \in m^i$ to mean the following: if $(\sigma_r^i, \boldsymbol{d}_{\sigma_r^i}) \to m^i_r$ is such that each $e \in m^i_r$ is contained in some edge $e \in m^i$, then we say $(\sigma_r^i, \boldsymbol{d}_{\sigma_r^i}) \in m^i$.

The proposal distribution we use at iteration $r$ is $p(\sigma_r, \boldsymbol{d}_{\sigma_r} | \theta^t)$, in which case the weight update for proposing $(\sigma^i_{r+1}, \boldsymbol{d}_{\sigma^i_{r+1}})$ from $(\sigma^i_r, \boldsymbol{d}_{\sigma^i_r})$ is:

\begin{align*}
	\alpha((\sigma^i_r, \boldsymbol{d}_{\sigma^i_r}) \to (\sigma^i_{r+1}, \boldsymbol{d}_{\sigma^i_{r+1}})) = 1[(\sigma^i_{r+1}, \boldsymbol{d}_{\sigma^i_{r+1}}) \in m^i].
\end{align*}

With the target and the proposal distribution clearly defined, we can sample the latent permutation and the decisions to approximate the Q function in Equation~(\ref{eq:Q-function}):

\begin{equation}\label{eq:approxQ}
	\tilde{Q}(\theta, \theta^t) = \sum\limits_{i=1}^{I} \frac{1}{N} \sum\limits_{n=1}^{N} \log L_c(m^i, \sigma^{i,n}, \boldsymbol{d}_{\sigma^{i,n}}) + \log p(\theta),
\end{equation}

\noindent where $(\sigma^{i,n}, \boldsymbol{d}_{\sigma^{i,n}}) \sim p(\sigma^{i}, \boldsymbol{d}_{\sigma^{i}} | \theta^t, m^i)$ for $n=1, ..., N$ for each $i = 1, ..., I$.

The M-step can be carried out using numerical optimization procedures. Since each decision is modelled using a multinomial logistic, the likelihood admits exact computation of the gradient. If the gradient can be computed exactly for $p(\theta)$, then efficient numerical optimization of the objective function over the parameters using off-the-shelf optimization routines such as L-BFGS [\cite{liu1989limited}] can be adopted. For example, if we take the isotropic Gaussian prior over $\theta$, then the objective function is:

\begin{align}\label{eq:objective-isotropic-gaussian}
	\tilde{Q}(\theta, \theta^t) = \sum\limits_{i=1}^{I} \frac{1}{N} \sum\limits_{n=1}^{N} \log L_c(m^i, \sigma^{i,n}, \boldsymbol{d}_{\sigma^{i,n}}) - \lambda \| \theta \|^2,
\end{align}

\noindent for some $\lambda > 0$.

\section{SMC Sampler for Prediction}
\label{sec:prediction}

In this section, we describe how to sample matchings, given the MAP estimate of the parameters (obtained by following the procedure described in Section~\ref{sec:inference}) and show how the samples can be used for prediction.  We base the exposition in this section on two important developments reported in the SMC literature. The first is the SMC samplers method [\cite{del2006sequential}], which extends basic SMC by introducing a sequence of intermediate distributions such that the SMC algorithm is defined on the common state space with the final distribution coinciding with the desired target distribution. This idea allows one to draw samples from an arbitrary state space. The second is the Poset SMC [\cite{Wang2015Bayesian}], which establishes conditions for developing SMC samplers for combinatorial state spaces. The version of the methodology presented here is more detailed than the one presented in \cite{jun2017sequential}.

\subsection{Background and Notation}

We begin by establishing notation for defining intermediate target distributions as well as intermediate state spaces. The state space of interest is the space of matchings on $G$, which we denote by $\mathcal{M}$. We generalize this space and introduce $\mathcal{M}_{r}, r = 1, ..., R$, as the intermediate state spaces. The state space $\mathcal{M}_r$ denotes a matching that can be realized after $r$ nodes have made decisions using our sequential decision model. We use $r$ to index the iterations of the SMC algorithm; therefore, $R$ is equal to the total number of nodes in the graph to be visited. This leads to $\mathcal{M}_{R} = \mathcal{M}$, the space where every node has made a decision and hence, placed into an edge. 

The intermediate distributions will be denoted by $\gamma_r$ and the proposal distribution by $\nu^+$. We will use $N$ to denote the number of particles in the SMC population and denote by $s_{r,n}$ and $w_{r,n}$ the particle $n$ at iteration $r$ and its un-normalized weight. The resampling step of the SMC algorithm is carried out using the normalized version of the weights, $\bar{w}_{r,n} = w_{r,n}/\sum_{n=1}^{N} w_{r,n}$. Note that the resampling step of the SMC algorithm induces the notion of a parent particle for each particle; we denote the index of the particle $s_{r, n}$ by $a_r^n$.  The weight computation is carried out in a recursive manner:
\begin{equation*}
	w_{r,n} = \bar{w}_{r-1,a_r^n} \times \alpha(s_{r-1,a_r^n}, s_{r,n}),
\end{equation*}
where $\bar{w}_{r-1,a_r^n} = 1/N$ if resampling is carried out in the previous iteration of SMC and $\alpha(s_{r-1,a_r^n}, s_{r,n})$ is the weight function:
\begin{equation*}
	\alpha(s_{r-1,a_r^n}, s_{r,n}) = \frac{\gamma_r(s_{r,n})}{\gamma_r(s_{r-1,a_r^n}) \nu^+(s_{r-1,a_r^n} \to s_{r,n})}.
\end{equation*}
The particles and weights at the final iteration are used to approximate expectations of functions $f : \mathcal{M} \to \R$ via
\begin{equation*}
	\E[f(M)] \approx \frac{1}{N} \sum\limits_{n=1}^{N} f(s_{R, n}),
\end{equation*}
if resampling is carried out after the last iteration and,
\begin{equation*}
	\E[f(M)] \approx \sum\limits_{n=1}^{N} \bar{w}_{R, n} f(s_{R, n}),
\end{equation*}
if resampling is not performed at the last iteration. We refer to \cite{doucet2009tutorial} for an excellent exposition of SMC methods.

\subsection{Partially Ordered Set}

An important notion that we need to introduce before we can complete the specification of our SMC sampler is that of a partially ordered set and how it arises in the SMC setting. 

A partial order $\prec$ defined on a set $\mathcal{S}$ is a binary relation that is reflexive, anti-symmetric, and transitive (often denoted as a pair, $(\mathcal{S}, \prec)$). The difference between a total order $<$ and the partial order $\prec$ is that not all elements of $\mathcal{S}$ are required to be comparable. That is, there may exist elements $s, s' \in \mathcal{S}$ such that neither $s \prec s'$ nor $s \succ s'$. The notion $s = s'$ for the partial orders is the same as for the total orders, i.e., $s = s'$ if and only if $s \prec s'$ and $s \succ s'$. We introduce the notion of Hasse diagram of a partially ordered set:

\begin{definition}
For $s, s' \in \mathcal{S}$, $s'$ \textbf{covers} $s$ if $s \prec s'$ and there does not exist $s'' \in \mathcal{S}$ such that $s \prec s'' \prec s'$.
\end{definition}

\begin{definition}
The Hasse diagram on $(\mathcal{S}, \prec)$ is an undirected graph $G=(\mathcal{S}, E)$ where the nodes of the graph are the elements of $\mathcal{S}$ and there is an edge between the nodes $s, s' \in \mathcal{S}$ if and only if $s$ covers $s'$.
\end{definition}

\begin{figure}[t!]
	\centering
	\begin{tabular}{cc}
		\includegraphics[width=0.6\textwidth]{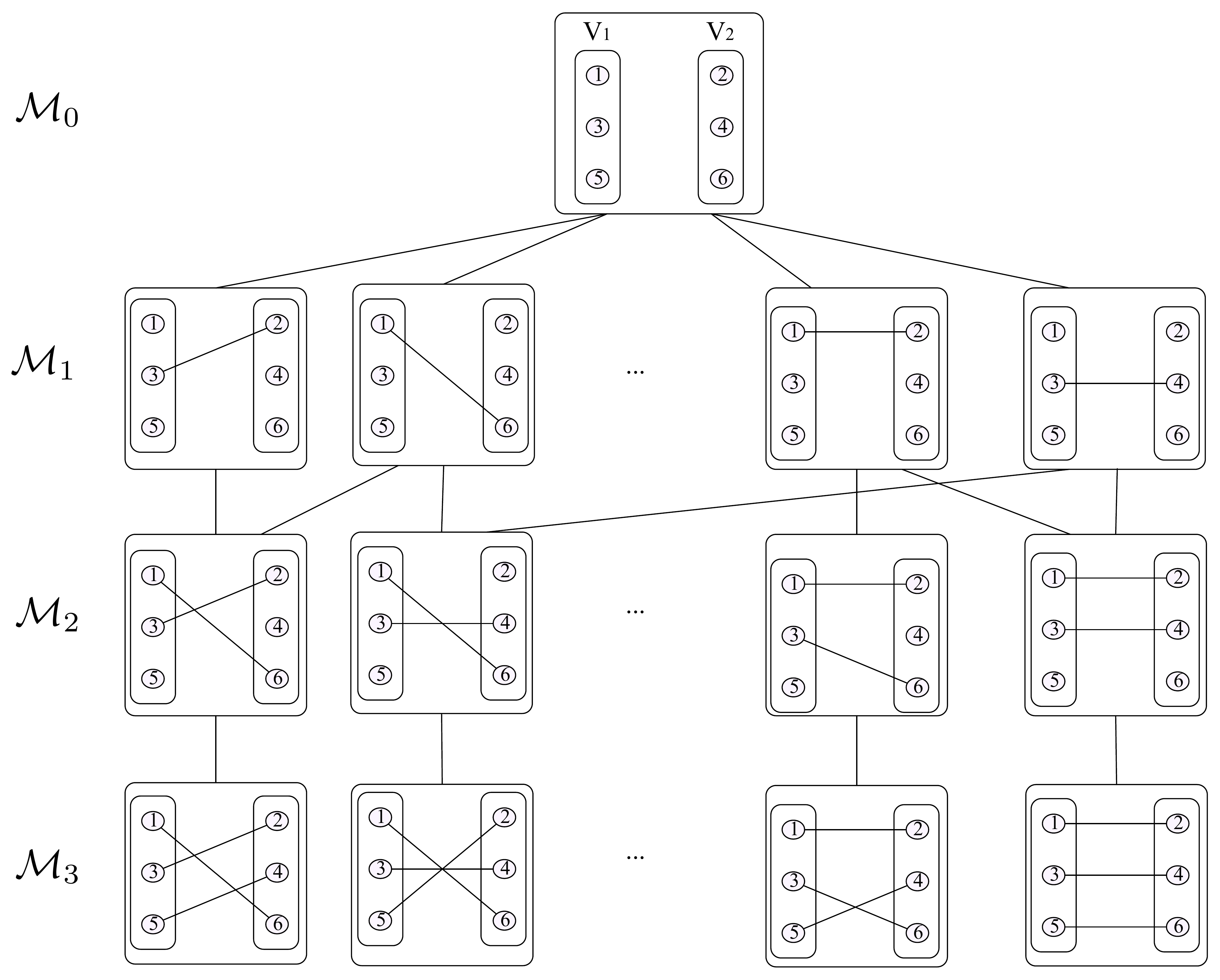} &
		\includegraphics[width=0.35\textwidth]{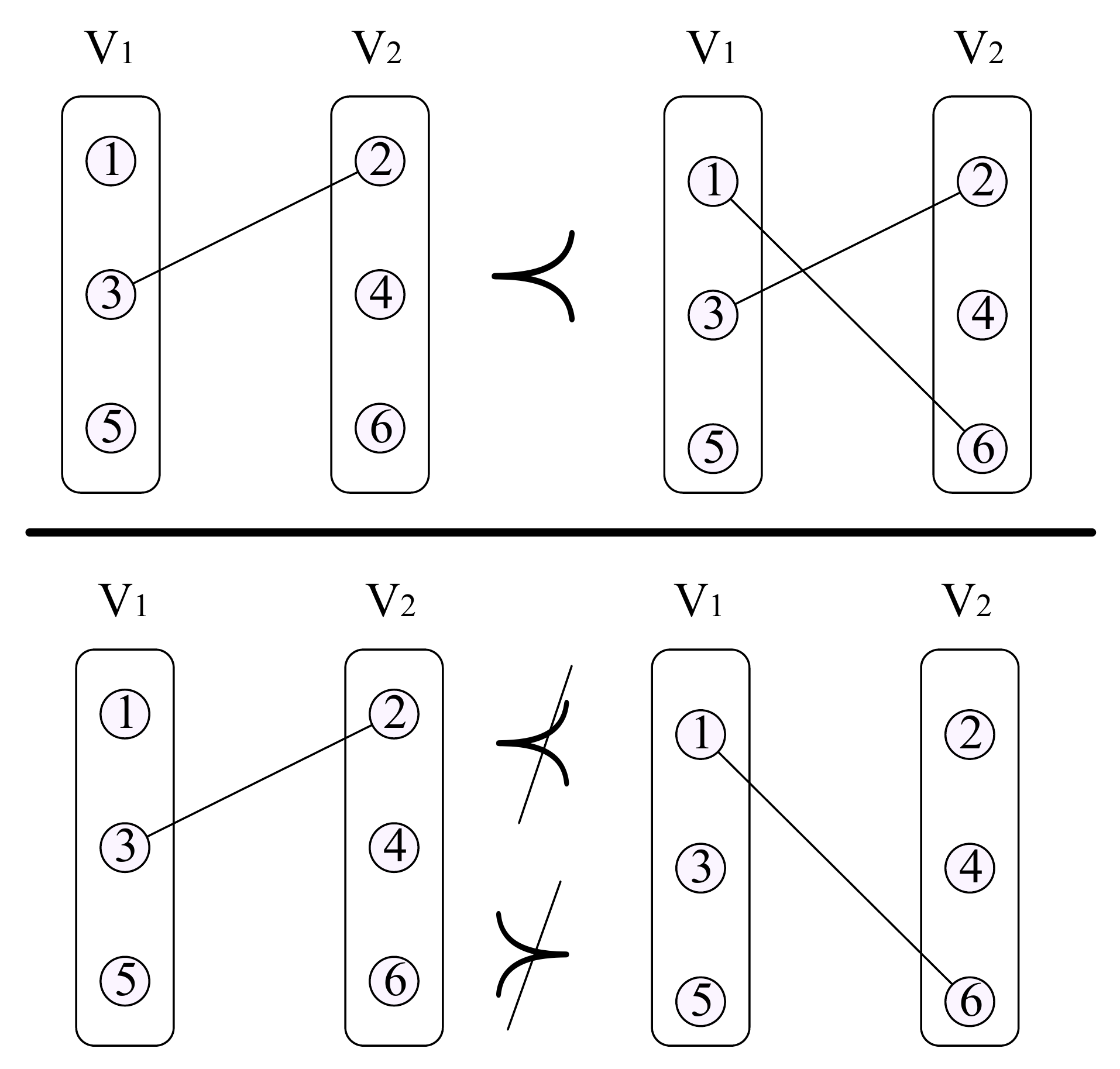} \\
		(a) & (b)
	\end{tabular}
	\caption{(a) Example of Hasse diagram corresponding to the bipartite decision model described in Section~\ref{sec:decision-model-bm}. (b) Example of partial order defined on bipartite matching. }
	\label{fig:partial-order}
\end{figure}

In the context of the sequential Monte Carlo sampler, the notion of a partial order on the state space $\mathcal{S}$ is characterized by the proposal distribution. In particular, $s$ covers $s'$ if $s'$ can be obtained by one application of the proposal to $s$. An example of a Hasse diagram corresponding to the decision model for the bipartite graph given in Section~\ref{sec:decision-model-bm} is shown in Figure~\ref{fig:partial-order} (a). We have also provided an example of a case where $s \prec s'$ on the top panel of Figure~\ref{fig:partial-order} (b) and a case where two states are not comparable in the bottom panel of Figure~\ref{fig:partial-order} (b). In essence, the sequential structure that is needed by the SMC method is induced by the partial order.

The next natural question concerns the conditions for a valid proposal distribution that ensures correctness of SMC algorithm. This is provided in [\cite{Wang2015Bayesian}] for combinatorial state spaces. One condition that is of great importance is that of \emph{connectedness}, that is, starting from an initial state $s_0$, one should be able to reach any other state $s \in \mathcal{S}$ by finite number of applications of the proposal distribution on $s_0$. 

\subsection{SMC Sampler for Graph Matching}
\label{sec:smcwsgm}

In Section~\ref{sec:inference}, we used SMC for sampling the latent variables $(\sigma, \boldsymbol{d})$. Our goal in this section differs in the sense that the object of interest includes matching as well as permutation and the decision sequences. In this section, we develop an SMC sampler that operates on an expanded state space that admits the sampling of matchings. First, recall that $(\sigma_r, \boldsymbol{d}_r)$ maps to a matching $m_r \in \mathcal{M}_r$. We define the intermediate state space as $\mathcal{S}_r = \mathcal{M}_r \times \Sigma_r \times \mathcal{D}_r$ and define the intermediate distribution as,

\begin{align}\label{eq:intermediate-distribution}
	\gamma_r(m_r, \sigma_r, \boldsymbol{d}_r | \theta) &= p(m_r | \sigma_r, \boldsymbol{d}_r) \times p(\sigma_r, \boldsymbol{d}_r | \theta) \nonumber \\
	&= 1[(\sigma_r, \boldsymbol{d}_r) \to m_r] \times p(\sigma_r, \boldsymbol{d}_r | \theta).
\end{align}

Here, $1[(\sigma_r, \boldsymbol{d}_r) \to m_r]$ denotes the indicator function that is $1$ if $(\sigma_r, \boldsymbol{d}_r)$ maps to $m_r$ and $0$ otherwise. 
Note that each $(\sigma_r, \boldsymbol{d}_r)$ maps to exactly one matching $m_r \in \mathcal{M}_r$ because each decision made by a knot results in it being placed in exactly one edge. 

The state space that our SMC sampler operates on is defined as $\mathcal{S} = \bigcup_r \mathcal{S}_r$. We can take the proposal distribution $\nu^+$ as the sequential decision model given in Equation~\ref{eq:local-model}. This choice ensures that the state space $\mathcal{S}$ is connected starting from the initial state $s_0 = (m_0, \sigma_0, \boldsymbol{d}_{\sigma_0})$, where $m_0 = \emptyset$. It is easy to verify that the weight function reduces to $1$ with this choice of the proposal for the intermediate distribution given in Equation~\ref{eq:intermediate-distribution}.

\subsection{Overcounting Correction}
\label{sec:overcounting}

\begin{figure}[t!]
	\centering
	\includegraphics[scale=0.3]{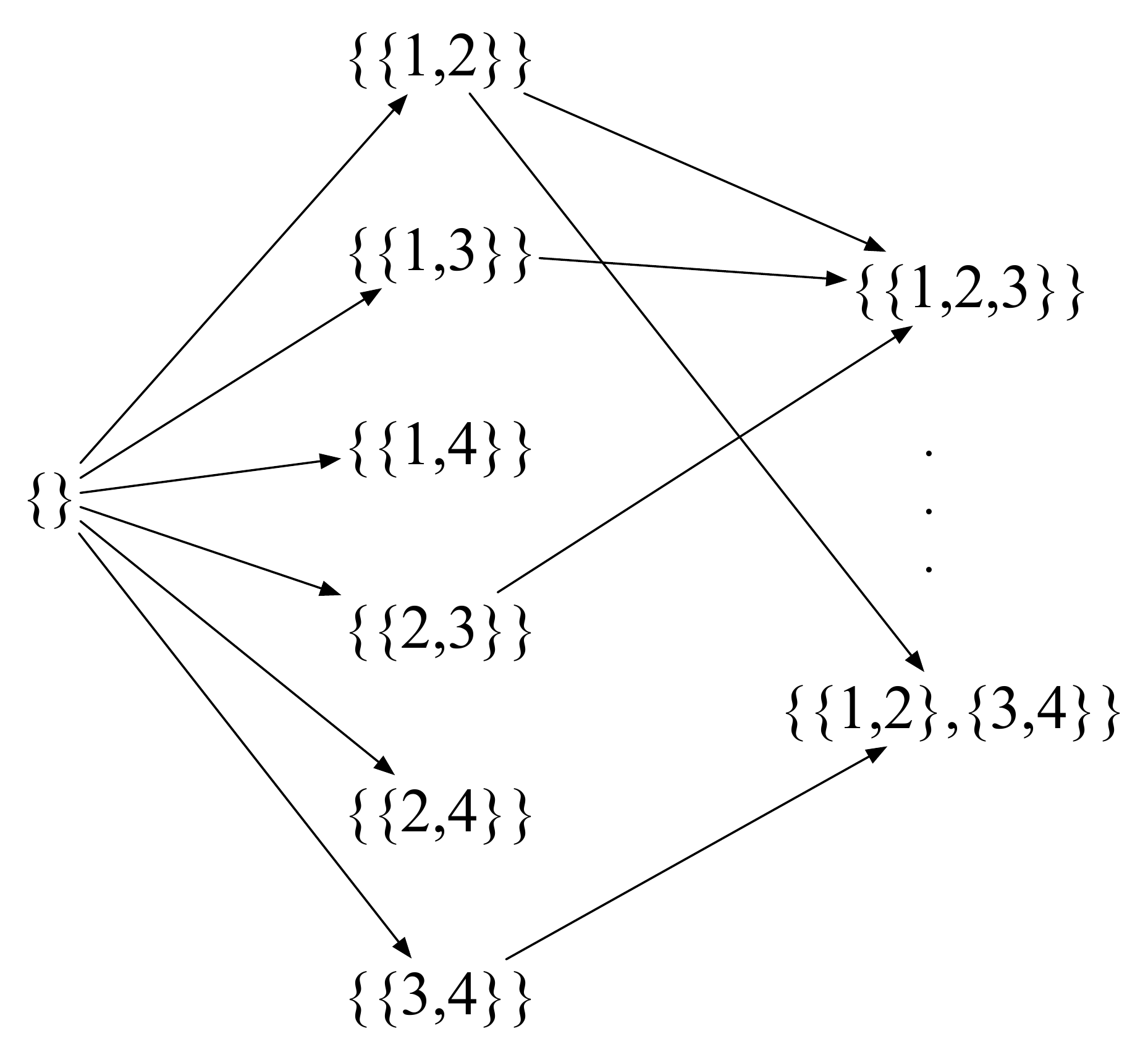}
	\caption{The state $\{\{1,2\}, \{3,4\}\}$ can be reached by two distinct paths from the initial state whereas the state $\{\{1,2,3\}\}$ can be reached by three distinct paths.}
	\label{fig:illustration-overcounting}
\end{figure}

Designing an SMC sampler for a combinatorial state space requires careful attention to the possibility of an overcounting problem, which may lead to biased estimates of the desired quantities [\cite{Wang2015Bayesian}]. The overcounting problem arises when there are multiple paths that lead to the same state. Consider a graph with 4 partitions and one node in each partition (see Figure~\ref{fig:simple-matching-example} (b)). The overcounting problem for this case is illustrated in Figure~\ref{fig:illustration-overcounting}. In this figure, we can see that there are 3 paths leading to the state $\{ \{1, 2, 3\} \}$ starting from the initial state whereas there are 2 paths leading to the state $\{ \{1, 2\}, \{3, 4\}\}$.  Approximation of any desired quantities using the SMC described in Section~\ref{sec:smcwsgm} would lead to bias if this overcounting problem were not corrected. A solution to this problem is to incorporate the backward kernel $\nu^{-}$ as proposed in [\cite{Wang2015Bayesian}]. One particular form of the backward kernel that works is,
\begin{equation}
	\nu^{-}(s' \to s) = |\mathcal{Q}(s')|^{-1} \times 1[\nu^+(s \to s') > 0],
\end{equation}
where $\mathcal{Q}(s')$ is the number of possible parent states of $s'$ and $\nu^+(s \to s') > 0$ if $s' \in \mathcal{S}$ can be proposed in one step starting from $s \in \mathcal{S}$. This leads to the weight computation step as,
\begin{align*}
\alpha(s_{r-1, a_r^n} \to s_{r, n}) &= \frac{\gamma_r(s_{r, n}) \times \nu^{-}(s_{r,n} \to s_{r-1, a_r^n})}{\gamma_{r-1}(s_{r-1, a_r^n}) \times \nu^+(s_{r-1, a_r^n} \to s_{r, n})} \\
&= \nu^{-}(s_{r,n} \to s_{r-1, a_r^n}).
\end{align*}
For the decision model corresponding to the knot matching application (see Section~\ref{sec:decision-model-knot-matching}), we have compiled a list of possible cases and the number of possible parents for each of the cases: 

\begin{itemize}
	\item If $s \in \mathcal{S}$ does not contain any singleton edge:
	\begin{itemize}
		\item For any edge with 2 nodes, if it contains
			\begin{enumerate}
				\item two visited nodes, count two possible parent states.
				\item one visited node, count one possible parent state.
			\end{enumerate}
		\item For any edge with 3 nodes, if it contains
			\begin{enumerate}
				\item two visited nodes, count two possible parent states.
				\item three visited nodes, count six possible parent states.
			\end{enumerate}
	\end{itemize}
	\item If there is at least one singleton edge in $s \in \mathcal{S}$:
	\begin{itemize}
		\item For any edge with 1 node, count one possible parent.
		\item For any edge with 2 nodes, if it contains
			\begin{enumerate}
				\item one visited node, this state is not reachable under this model.
				\item two visited nodes, then count two possible parent states.
			\end{enumerate}
		\item For any edge with 3 nodes, if it contains
			\begin{enumerate}
				\item two visited nodes, then return zero possible parent state.
				\item three visited nodes, then count three possible parent states.
			\end{enumerate}
	\end{itemize}
\end{itemize}

\begin{figure}[t!]
	\centering
	\begin{tabular}{ccc}
		\includegraphics[scale=0.3]{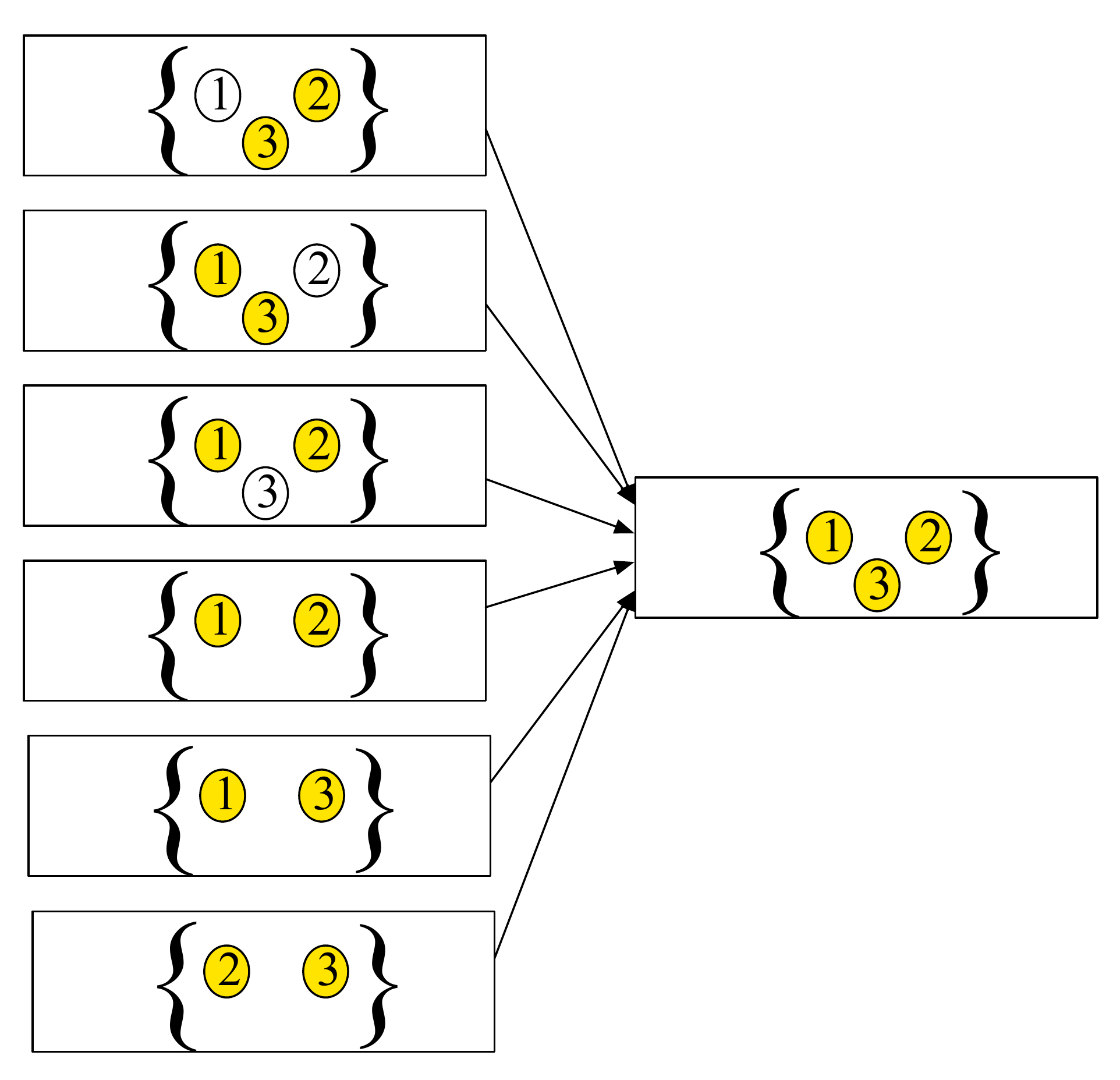} &
		\includegraphics[scale=0.3]{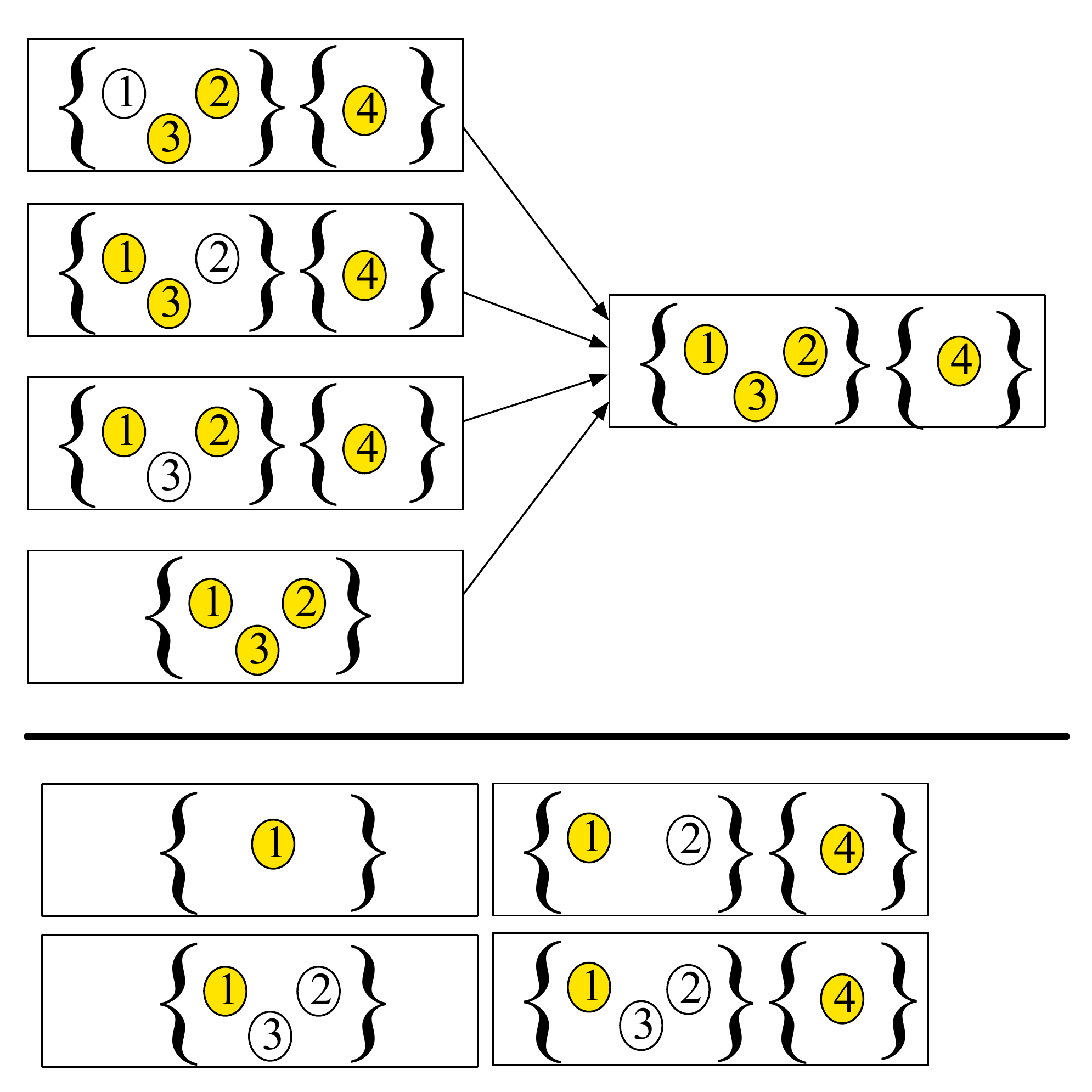} \\
	\end{tabular}
	\caption{The possible parent states for different cases. Left: a state containing a 3-matching where all three nodes have been visited. Top right: containing a 3-matching and a singleton where all four nodes have been visited. Bottom right: example of states that are not permitted under the decision model for the knot matching application (see Section~\ref{sec:decision-model-knot-matching}).}
	\label{fig:parent-particles}	
\end{figure}

Note that if a singleton set exists in a state, and if there are 2 visited nodes in a 3-matching, then undoing the move performed by one of the visited nodes in the 3-matching breaks it into 2-matching, which produces a state where a singleton cannot have been attained. Hence, the last move must have been made by one of the singletons, which means there can only be one parent state (which is accounted for by the singleton edge). Figure~\ref{fig:parent-particles} is an illustration of different cases.

\subsection{Evaluation Metrics}
\label{sec:evaluation-metric}

In this section, we describe two approaches to evaluating the goodness of the population of the particles generated using the SMC sampler for matching.

\subsubsection{Single Sample Prediction}

We can obtain a single sample prediction by choosing the particle with the highest likelihood, denoted $\hat{m}$. We can then compute the prediction accuracy as,

\begin{equation}
	a(\hat{m}, m_{true}) = \frac{1}{|m_{true}|} \sum\limits_{e \in m_{true}} 1[e \in \hat{m}],
\end{equation}

\noindent where $m_{true}$ denotes the true matching. Note that $a(\hat{m}, m_{true}) \in [0, 1]$ and it is equal to 1 when $\hat{m} = m_{true}$.

\subsubsection{Jaccard Index}

To assess the entire particle population, we use Jaccard index, which is commonly used metric for computing the similarity coefficient [\cite{levandowsky1971distance}]. We can compute the Jaccard index to evaluate the deviation of each of the SMC particles from the true matching. Jaccard index is defined on two sets $A$ and $B$ as follows:

\begin{equation}\label{eq:jaccard-idnex}
  	J(A, B) = \frac{| A \bigcap B|}{|A \bigcup B|}.
\end{equation}

We can use Jaccard index to evaluate a particle $m_n$ as follows. For each node $v$, we find the edge that contains $v$ in $m_n$ as well as in $m_{true}$. We will denote these edges by $m_n(v)$ and $m_{true}(v)$. Then, we compute the Jaccard index between the $n$-th particle and the truth by,

\begin{equation}
	J(m_n, m_{true}) = \frac{1}{|V|} \sum\limits_{v \in V} J(m_n(v), m_{true}(v)).
\end{equation}

Note that the minimum value for $J(m_n(v), m_{true}(v))$ is $1/6$ for the knot matching application since both $m_n(v)$ and $m_{true}(v)$ must contain $v$. The maximum value is $1$ if the two edges contain the same set of nodes.

%% file: simulation.tex
\section{Generating synthetic data}
\label{sec:simulation}

This section presents a mechanism to simulate synthetic boards that closely mimics the real data. This simulation mechanism is needed partly due to the prohibitive cost associated with obtaining the real data, which limits the study of knot formulation and implementation of new covariates to accurately capture the variety of knot shapes.


We first conceptualize tree branches as approximately cone-shaped objects emanating from the center of the tree trunk [\cite{guindos2013three}]. As the tree is cut into construction lumber, these cones intersect with rectangular prisms representing the pieces of lumber, forming knot faces on the board's surfaces. Thus synthetic boards and knot faces with known matchings and realistic geometry can be generated by simulating locations and sizes of cones representing the tree branches, and calculating the conic sections with the four planes representing the surfaces of the board. Conic sections arising from the same branch are matched knot faces.

The board is situated in 3-D Cartesian coordinates as described in Section \ref{sec:background-data}, with length 5000 units ($x$ dimension), width 300 units ($y$ dimension) and height 150 units ($z$ dimension). Let $n_k$ denote the random number of knots on the board.  Based on the number of knots observed in real data, we draw $n_k \sim \text{Poisson}(\rho)$ and generate $n_k$ branches that intersect with the board. Lumber is cut so that most branches go through the two `wide' surfaces, so we initially position a branch according to the equation of a right circular cone that opens upward from the origin,
$$
\frac{x^2 + y^2}{c_0^2} = z^2,
$$
where the random $c_0$ governs the slope of the cone and we restrict $z>0$. It is possible for branches emanating in different directions to appear in one board, so with probability 1/2 we allow the cone to open downwards by reflecting it over the plane $z = 75$. We next apply a random angle of rotation to the cone around the $x$ and $y$ axes, to mimic the variability in the angles of tree branches. Then, the center of the cone is translated to a random $(x,y)$ location on the board. Additional variation in the sizes of knot faces is provided by a random translation in the $z$ direction. Intersections of the final cone and the four surfaces are computed using a numerical root-finding procedure, and ellipses are fitted to the intersections representing the knot faces.

The cone simulation is repeated for each of the $n_k$ branches required. Since different tree branches do not intersect, we impose a condition to reject simulated cones that overlap geometrically or are otherwise too close to existing cones.  Consider the 3-D line segment joining the centers of two knot faces due to the same branch. Then, the line segments corresponding to different branches cannot be too close: specifically, the minimum distance $d$ between the two line segments should exceed the typical diameter of a branch. Hence, a simulated cone is rejected and resampled if $d < 50$ with an existing cone.  This corresponds to a real distance cutoff of about 0.6in.

The procedure for generating a board is summarized in Algorithm \ref{algo:synthetic-board} with the specific simulation parameters used.  We use this procedure to generate the samples of synthetic boards used in the computational experiments in the following section.

\begin{algorithm}
\begin{algorithmic}[1]
\STATE Draw $n_k \sim Pois(25)$
\FOR{$i = 1, \ldots, n_k$}
	\STATE Draw cone parameters: slope $c_0 \sim Unif[0.025,0.05]$, orientation $s \sim Bern(0.5)$ \label{algo:startcone}
	\STATE Draw rotation angles: $\theta_x, \theta_y \stackrel{iid}{\sim} Unif[ -\pi/6, \pi/6]$
    \STATE Draw center $(x_t,y_t)$: $x_t \sim Unif[0,5000]$, $y_t \sim Unif[0,300]$
    \STATE Draw $z$ translation: $z_t \sim (2s-1) Unif[0,500]$
    \FOR{$j = 1,2,3,4$}
        \IF{cone intersects with surface $j$}
            \STATE Compute center and covariance matrix for ellipse of conic section on surface $j$
            \STATE $t_{ij} \leftarrow 1$
        \ELSE
            \STATE $t_{ij} \leftarrow 0$
        \ENDIF
    \ENDFOR
    \STATE Compute line segment $L_i$ between ellipse centers on two surfaces with $t_{ij}=1$
    \FOR{$b = 1, \ldots, i-1$}
        \STATE $d_{i,b} \leftarrow $ minimum distance between $L_i$ and $L_b$
        \IF{ $d_{i,b}<50$}
            \STATE \textbf{goto} \ref{algo:startcone}
        \ENDIF
    \ENDFOR
    \STATE Define matching $m_i = \{ j: t_{ij} = 1 \}$
\ENDFOR
\end{algorithmic}
\caption{: {\bf Synthetic board generator}}
\label{algo:synthetic-board}
\end{algorithm}

%% file: 5data-analysis.tex

\section{Experimental Results}
\label{sec:application}

This section presents the results of experiments to demonstrate the performance of the methods proposed in this paper. We have 30 boards that were manually annotated (i.e., knots were matched manually) for evaluation purposes. As it is expensive to acquire additional data, we used the procedure described in Section \ref{sec:simulation} to simulate additional pieces of lumber to supplement the real data for analysis, in particular to test the feasibility of deploying the methodology under the real time constraint.

\subsection{Preliminary Experiments}

Before tackling the knot matching data, we perform experiments to validate various components of the model and the methods proposed in this paper.

\begin{figure}[t!]
	\centering
	\begin{tabular}{c}
		\includegraphics[scale=0.95]{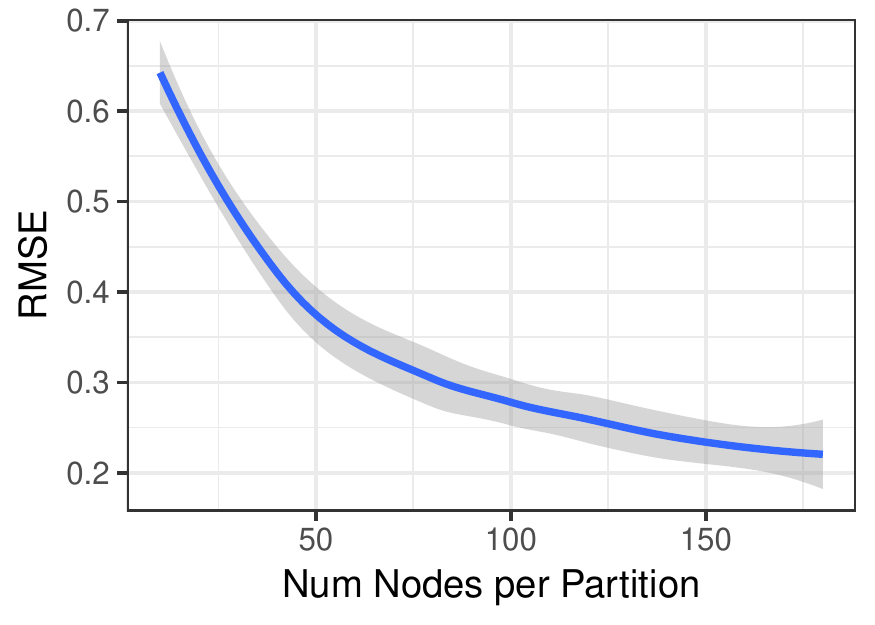} \\
		(a) \\
		\includegraphics[scale=0.55]{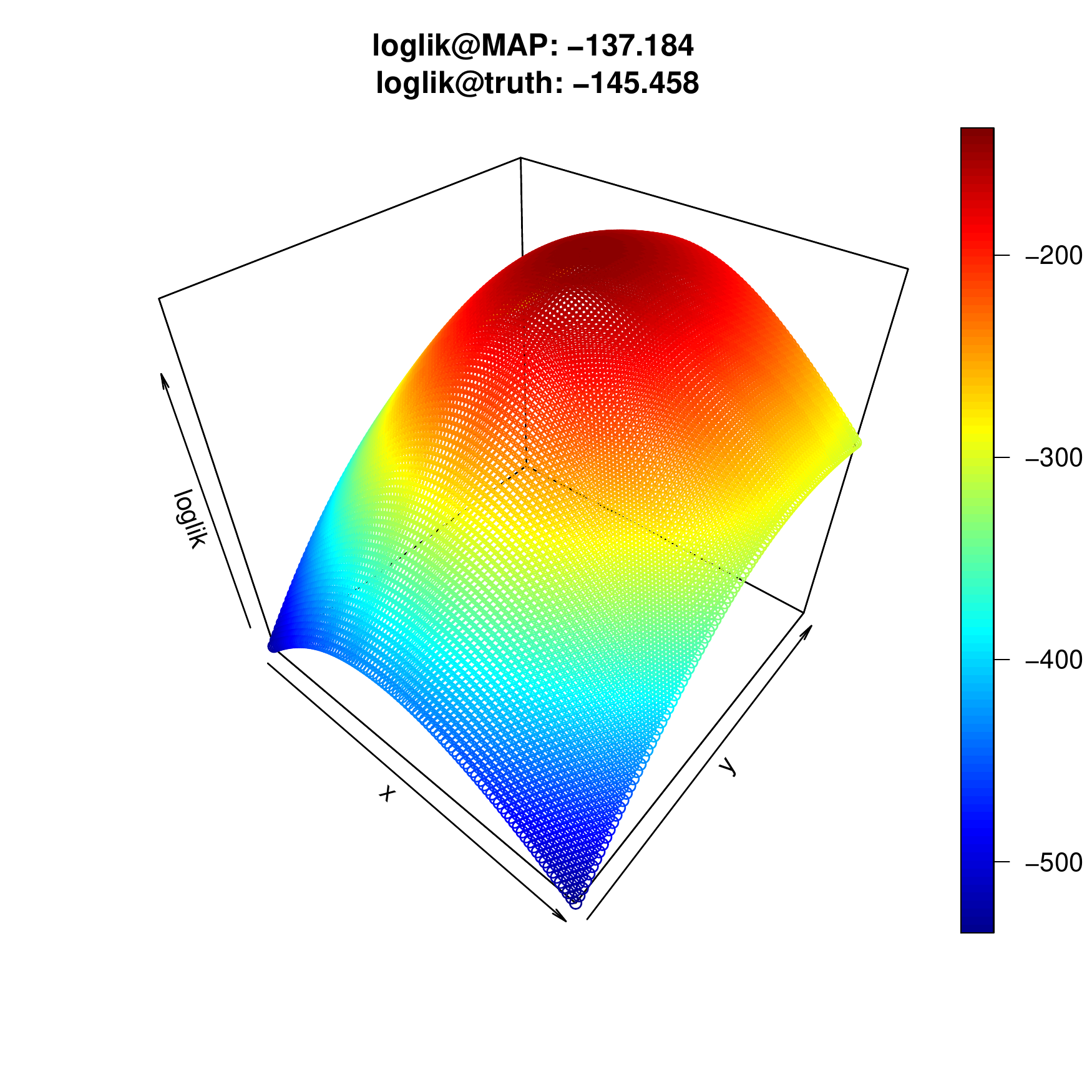} \\
		(b)
	\end{tabular}
	\caption{Experiments where the sequence $\sigma$ is given. (a) The plot of RMSE as the number of nodes is increased. (b) The sample posterior surface showing that the MAP estimate correctly finds the mode of the posterior.}
	\label{fig:map_known_sigma}
\end{figure}

\subsubsection{Validation of Parameter Estimation Procedure}
\label{sec:exp-param-estimation}

We begin with a simple parameter estimation experiments. We simulate a synthetic graph as follows:

\begin{enumerate}
	\item Sample the parameters, $\theta_j \sim N(0, \tau^2)$.
	\item Generate $N$ nodes per partition.
	\item For each node, sample the covariates $f_j \sim N(0, \zeta^2)$ for $j=1,...,d$, where $d$ denotes the number of covariates.
	\item Sample $\sigma$ from uniform distribution over the permutation.
	\item Sample the decisions $d_{\sigma} \sim p(\cdot|\sigma, \theta)$ (using Equation~\ref{eq:local-model}).
\end{enumerate}

We can repeat the synthetic graph generation process $I$ times to obtain a set of labelled matchings $\{(m^i, \sigma^i, \boldsymbol{d}_{\sigma^i})\}_{i=1}^{I}$. Given this dataset, we carried out experiments to validate the MAP estimator of the parameters. In this experiment, we used the true $\sigma$ that was used for generating each of the matchings. In this case, we expect the parameter estimator to be efficient as it boils down to that of estimating the parameters of a model composed of multiple multinomial logistic regression models. We have shown the root mean squared error, $d^{-1} \|\hat{\theta}_{MAP} - \theta_{true}\|_2$, when $d=2$ in Figure~\ref{fig:map_known_sigma} (a). As the number of nodes in the graph is increased, the accuracy of the estimation improves as expected.  We have also generated a surface plot of the posterior when $d=2$ (two covariates) in Figure~\ref{fig:map_known_sigma} (b). Note that the response surface that we are optimizing over is convex and the MAP estimate attains a higher value of the log-likelihood compared to the truth. Figure~\ref{fig:map_known_sigma} (a) was obtained using $I=10$. We fixed the number of partitions to $2$. The standard error estimates were obtained using LOESS in the R package \emph{ggplot2} [\cite{ggplot2}]. Figure~\ref{fig:map_known_sigma} (b) was obtained by evaluating the likelihood function over a grid of parameter values. This experiment serves to verify the correctness of our parameter estimation procedure.



\subsubsection{Overcounting Correction Experiments}

\begin{figure}[t!]
	\centering
	\begin{tabular}{cc}
		\includegraphics[width=2cm,height=5cm]{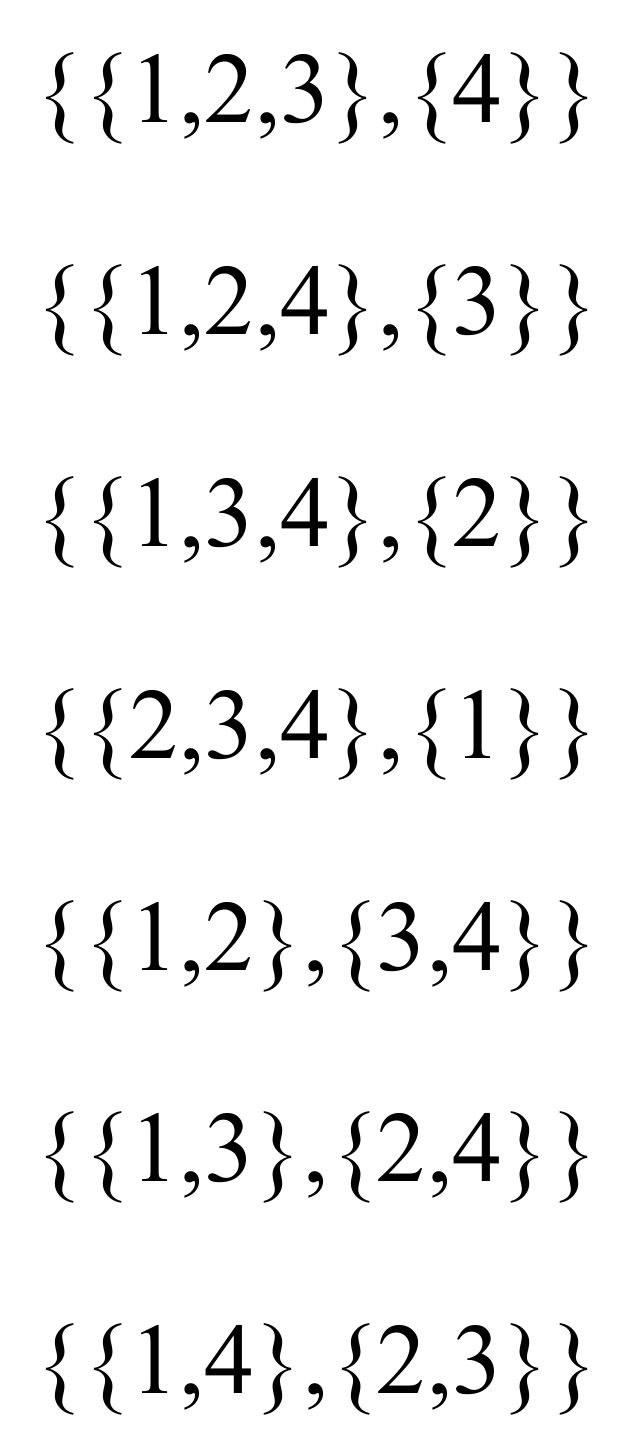} &
		\includegraphics[scale=0.8]{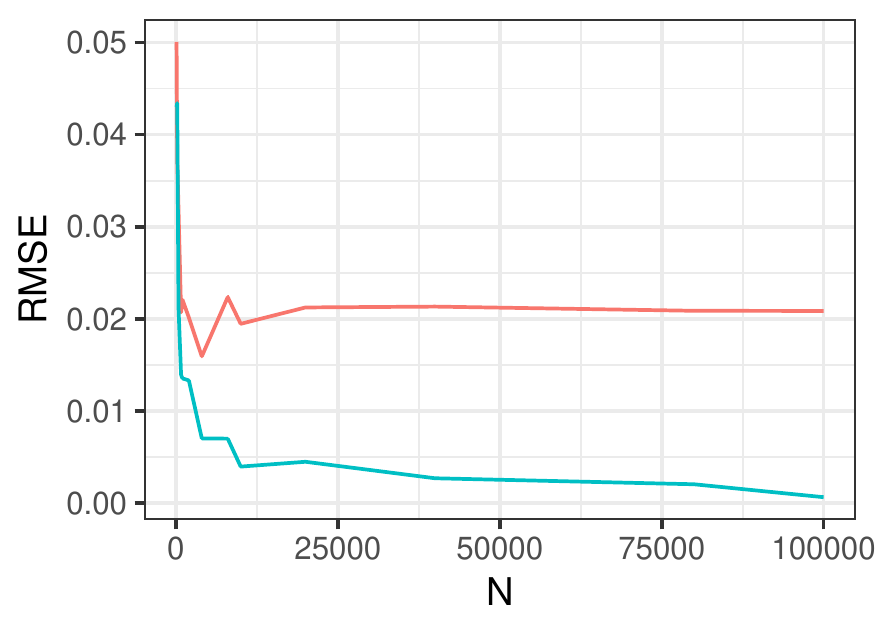} \\
		(a) & (b)
	\end{tabular}
	\caption{Overcounting problem illustrated on sampling from uniform graph matching. (a) The number of possible $\{2, 3\}$-matchings for a graph with four partitions and one node in each partition. (b)The root mean squared error with overcounting correction (blue), without the overcounting correction (red). }
	\label{fig:overcounting-experiments}
\end{figure}

This subsection illustrates the overcounting problem and why it needs to be addressed to sample graph matching using SMC. To that end, we assume a scenario where we want to sample graph matching from the uniform distribution over all possible configurations permitted by our choice of the decision model (for example, the decision model for knot matching given in Section~\ref{sec:decision-model-knot-matching}). For illustrative purposes, suppose we have a simple example of a quadripartite graph with one node in each partition. Note that this decision model is restricted to $\{2, 3\}$-matchings so there are total of 7 matchings possible for this graph and hence, we expect the probability of sampling a matching to be $1/7$ (see Figure~\ref{fig:overcounting-experiments} (a)).

We have computed the estimate of the probability of each matching configuration from the SMC population, $\hat{p}_m$ and computed the root mean squared error: $\sqrt{7^{-1} \sum_{m \in \mathcal{M}} (\hat{p}_m - 1/7)^2}$. In Figure~\ref{fig:overcounting-experiments} (b), we show that the RMSE tends to $0$ as the number of particles used in the SMC is increased (the blue curve). On the other hand, the RMSE stabilizes around 0.02 when the overcounting problem is ignored (the red curve).

\subsection{Data Analysis}

In this section, we analyze the simulated and real lumber data.

\subsubsection{Real Data Analysis}
\label{sec:real-data-analysis}

\begin{figure}[t!]
	\centering
	\includegraphics[width=0.85\textwidth]{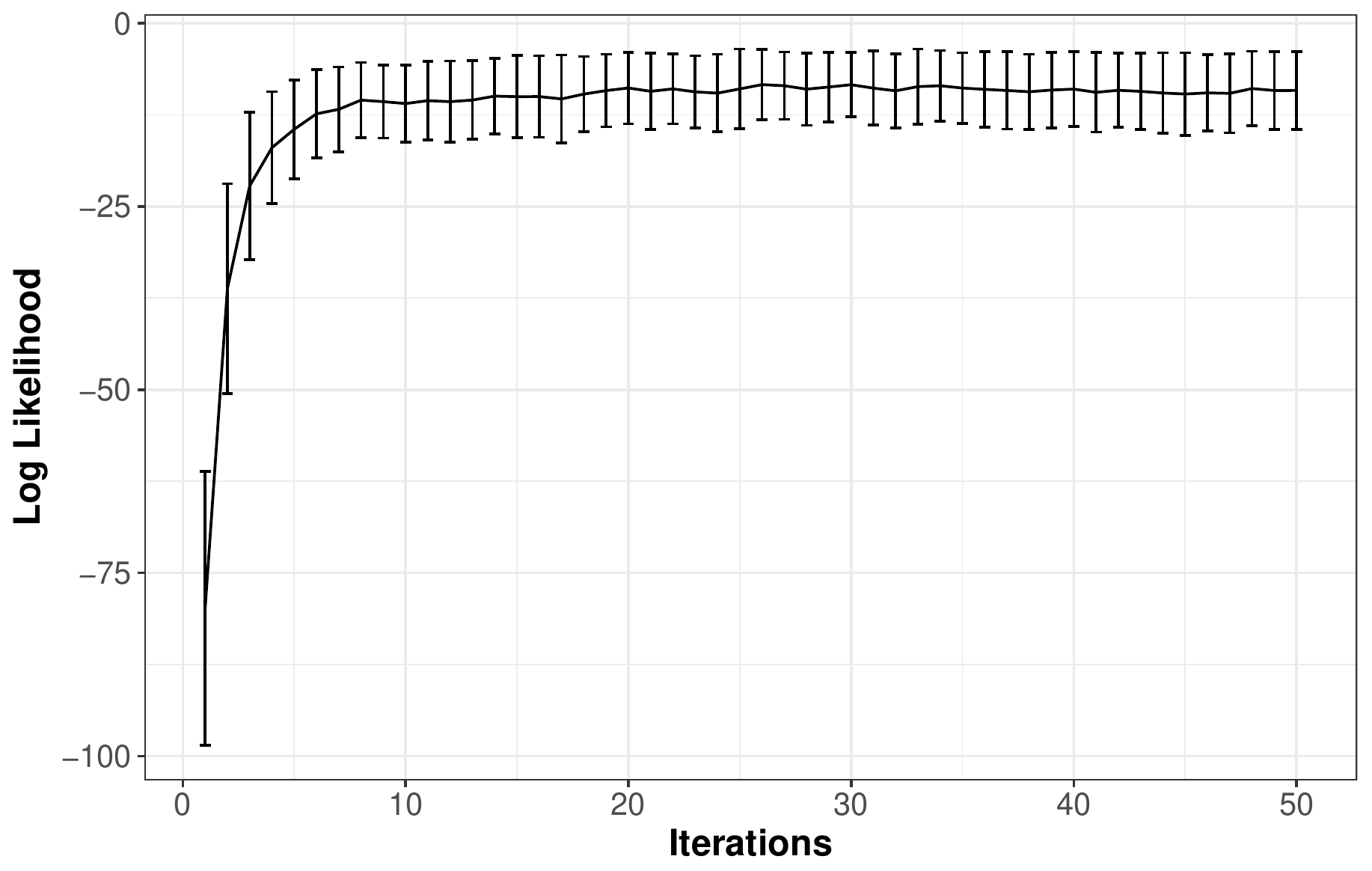}
	\caption{The plot of the $\tilde{Q}$ function versus iterations. The error bars correspond to $\tilde{Q}$ plus/minus two times the sample standard deviation (i.e., $\tilde{Q} \pm 2 \hat{\sigma})$.  The convergence of MC-EM seems to have been reached in 10 iterations.}
	\label{fig:mcem-diagnostics1}
\end{figure}

\begin{figure}[t!]
	\includegraphics[scale=0.7]{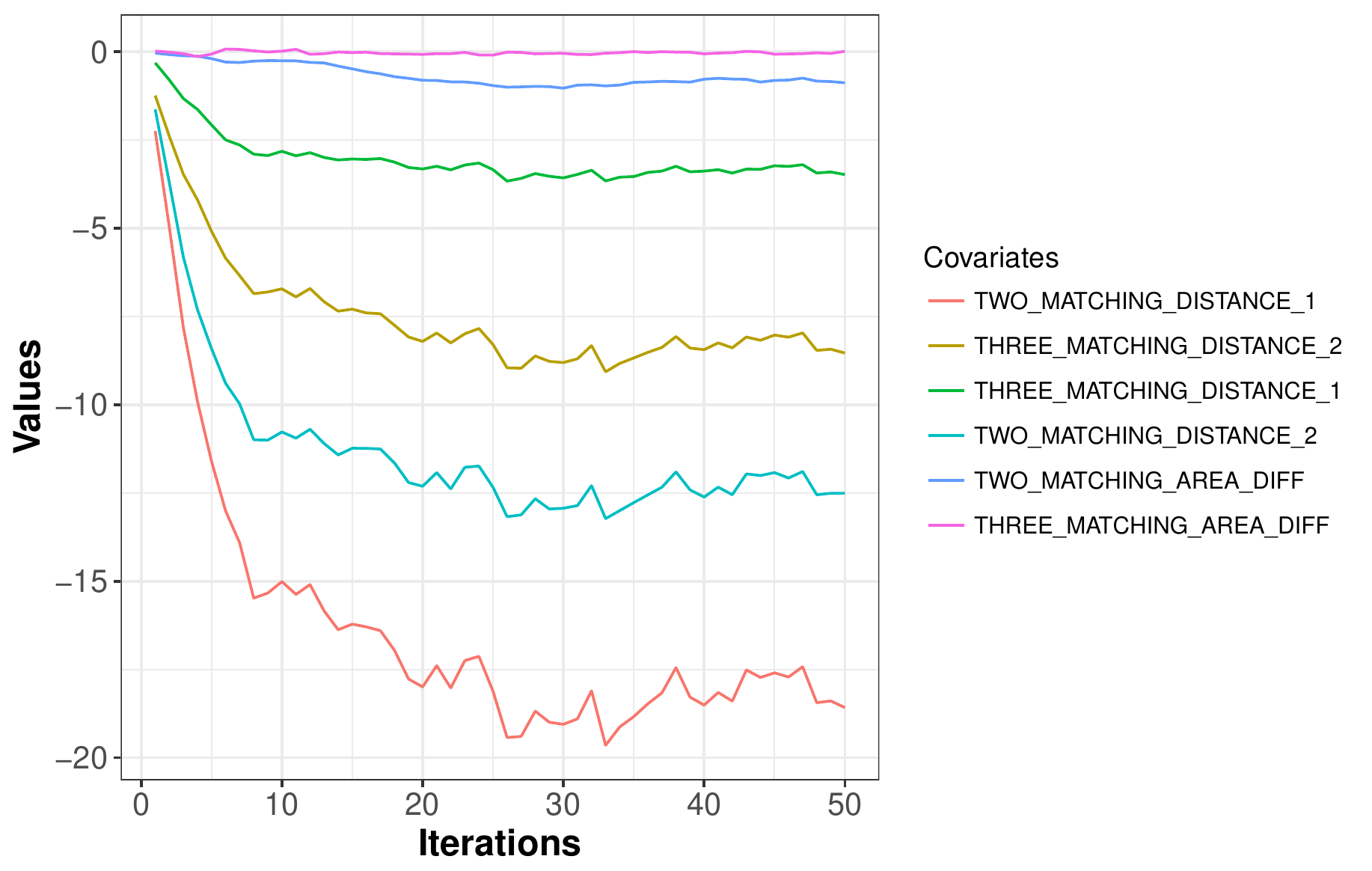}	
	\caption{The trajectory of parameters in the MC-EM training versus iterations. The distance based covariates seem to play important roles in determining the correct matches compared to the area based covariates.}
	\label{fig:mcem-diagnostics2}
\end{figure}

In this section, we evaluate our methods on the 30 boards that had been manually annotated. First, we illustrate the parameter estimation procedure that was carried out using an MC-EM procedure. The sample size used for the E-step is kept at 100 for the first 10 iterations of MC-EM, which is increased to 500 onwards to reduce the Monte Carlo error across the iterations. The convergence of MC-EM is monitored by plotting $\tilde{Q}$ across the iterations. In Figure~\ref{fig:mcem-diagnostics1}, we show a plot of the $\tilde{Q}$ function across MC-EM iterations with the error bars computed using the standard deviation of the Monte Carlo samples to $\tilde{Q}$ at each iteration of MC-EM. The figure suggests that convergence is reached in about 10 iterations. The trajectory of the parameters across MC-EM iterations is shown in Figure~\ref{fig:mcem-diagnostics2}. This plot shows that distance based covariates play important roles compared to area based covariates. The value for $\lambda$ is set to $1$ for the experiments. 

\begin{figure}[t]
	\includegraphics[width=0.95\textwidth]{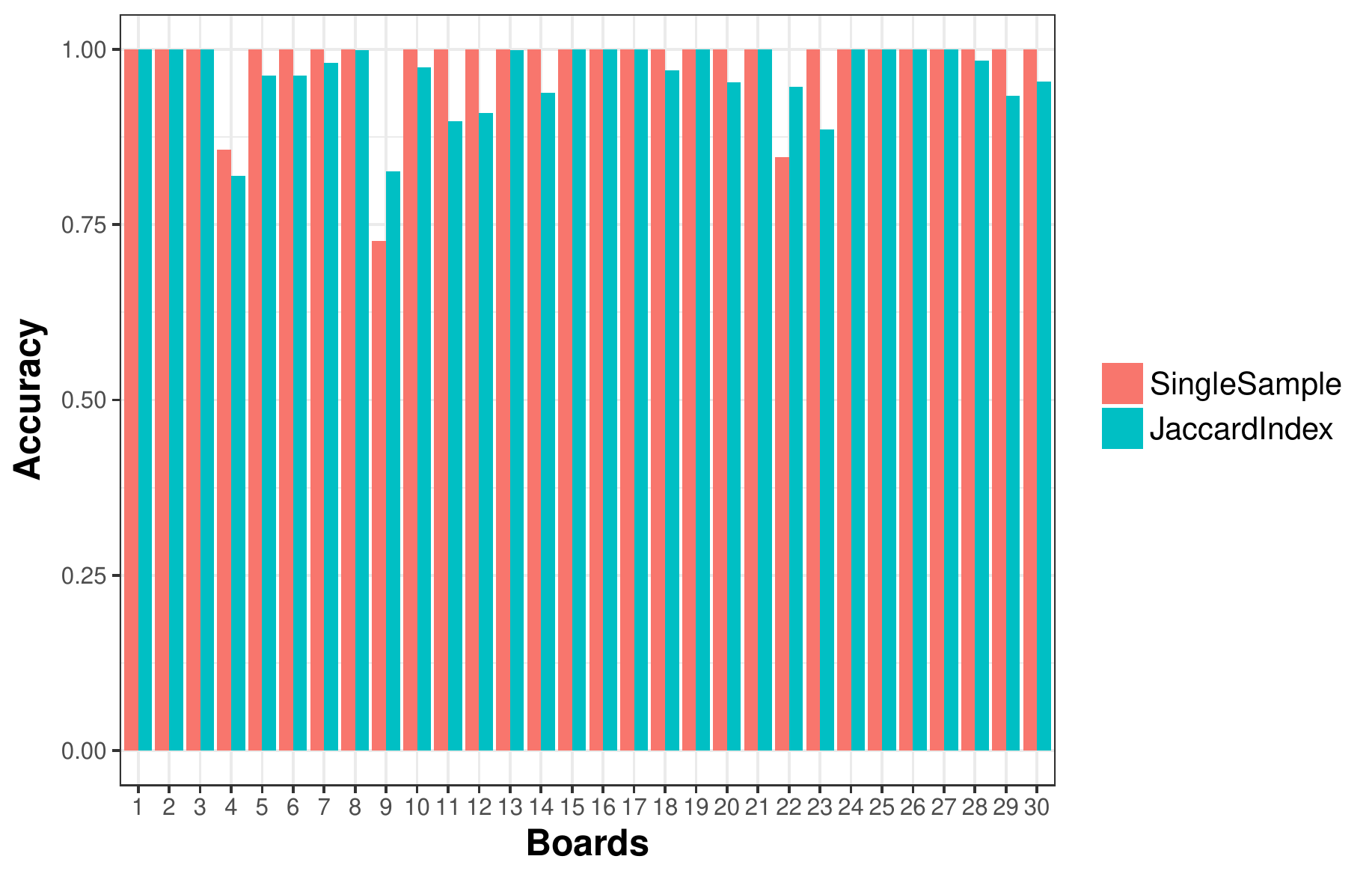}
 	\caption{Single sample prediction accuracy and Jaccard index evaluation of the matching samples generated by SMC computed on the real data for each board. The single sample prediction accuracy is perfect for all but 3 boards. The quality of the matchings generated for each board by SMC appears reasonable based on the values of Jaccard index.}
 	\label{fig:real-data-results}
\end{figure}

To evaluate predictive performance, we perform leave-one-out cross validation. That is, we leave one board out from the MC-EM inference procedure (i.e., obtain MAP estimate using the remaining 29 boards). Then, we sample graph matchings on the held-out board to be evaluated using single sample prediction accuracy and the Jaccard index described in Section~\ref{sec:evaluation-metric}. With $\lambda = 1$, the overall accuracy is $375/384 = 0.977$ using a single sample prediction. The board-by-board performance is shown in Figure~\ref{fig:real-data-results}.  We have experimented with $\lambda = 0.1$ and $\lambda = 10$ as well and found the single sample prediction performance to be comparable to $\lambda = 1$.

\subsubsection{Simulated Data}

A data simulation procedure is helpful for future research in the automatic strength grading of lumber as it can be used to calibrate the performance of matching methodology presented in the paper on simulated boards due to the high cost associated with acquisition of the real data (i.e.,~requiring manual knot matching). In particular, there are various types of knots that we have not been able to model due to the limitation in the dataset. The simulation provides a testbed to develop new models for knots and new features for knot matching for the application experts.

One way that we make use of the simulated data is to test the feasibility of deploying the SMC sampler in real time. As the end goal is to deploy the matching mechanism developed here to mills that operate under real time constraints, it is important to study the time it takes to generate samples using SMC. To that end, we simulated 100 boards. To speed up the sampling procedure, we segmented each board into multiple subgraphs. This segmentation procedure was carried out based on distance so that knots within certain distances are placed into the same subgraph. The SMC sampler was executed locally within each subgraph and this helped to significantly reduce the time to draw samples since there were less decisions to consider at each iteration of an SMC. 

\begin{figure}[t!]
	\includegraphics[width=0.95\textwidth]{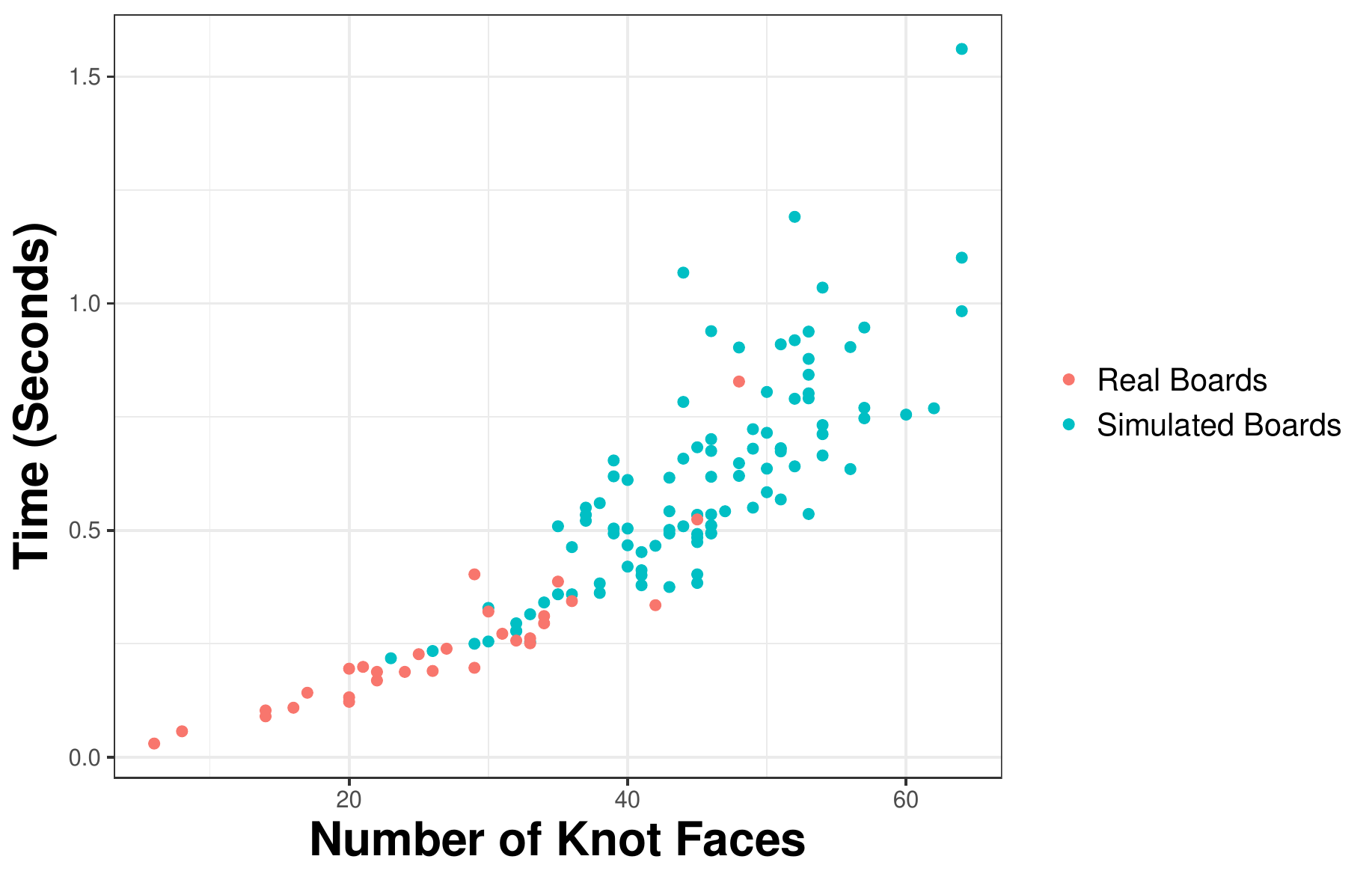}
	\caption{Timing results on the simulated and the real boards. The prediction times are within a second for the real boards. The timing results on the simulated boards serve to test the feasibility of deploying the SMC sampler in lumber mills.}
	\label{fig:timing-results}
\end{figure}

We have plotted the timing results in Figure~\ref{fig:timing-results}. The figure depicts the scatter plot of the timing results for 100 simulated boards as well as 30 real boards when the number of particles is set to $1000$ against the number of knot faces on the board. Observe that for most boards, it only takes a fraction of a second for sampling to complete. For completeness, we carried out a 2-fold validation procedure to quantify the performance of our methodologies on the simulated dataset. We attained the single sample prediction accuracy of 93\% on the 100 simulated boards.

%% file: 6conclusion.tex
\section{Conclusion and Discussion}
\label{sec:conclusion}

In this paper, we propose a statistical inference procedure for matching on $K$-partite non-uniform hypergraph. We developed methods for the novel knot matching application, which can be formulated as a 4-partite hypergraph matching problem. This is an important step towards automating the grading of lumber, one where statistical inferential methods can be used to produce not just a single strength prediction value but a posterior predictive interval that captures any uncertainties encountered in the process.  We developed a sequential decision model that admits efficient inference of the parameters and an SMC sampler that allows for rapid sampling that can meet the real time constraints of lumber mills.  Furthermore, the model could in principle admit a full Bayesian approach via particle MCMC methodology [\cite{andrieu2010particle}]. 

The predictive performance of our methodology will be dependent on the choice of covariates. The framework we have laid out in this paper is general and will allow users to craft and experiment with different covariates, especially when more sample data becomes available in the future. The covariates that we crafted here led to good performance on the current sample data; however, they are by no means complete. For example, additional information that might be incorporated to further improve predictive performance includes the rotation angle of the knot faces and/or more detailed shape information on the knots. 

Future work remains to complete an automatic lumber strength grading pipeline. We shall develop an enhanced statistical model for strength prediction, using the output from our knot matching methodology as an input for producing the strength estimate. With accurate knot matchings and uncertainty appropriately quantified, we anticipate that our contributions will have practical impact in this discipline and demonstrate yet another application where modern statistical methods can greatly benefit the field.